\documentclass[prd,aps,twocolumn,preprintnumbers,amsmath,amssymb]{revtex4}
\pdfoutput=1
\usepackage{amsmath} 
\usepackage{amssymb} 
 
\usepackage{nonfloat}
 
\usepackage{graphicx}
\usepackage{dcolumn}
\usepackage{bm}
\def\be{\begin{equation}} 
\def\ee{\end{equation}} 
\def\bea{\begin{eqnarray}} 
\def\eea{\end{eqnarray}} 

\newcommand{\comment}[1]{}

\begin{document} 
 
 
\date{\today} 
 
\title{Searching for Signatures of Cosmic String Wakes in 21cm Redshift Surveys 
using Minkowski Functionals} 
 
\author{Evan McDonough and Robert H. Brandenberger} 

\affiliation{Department of Physics, McGill University, Montr\'eal, QC, H3A 2T8, Canada}

 
\begin{abstract} 

Minkowski Functionals are a powerful tool for analyzing large scale structure, in particular 
if the distribution of matter is highly non-Gaussian, as it is in models in which
cosmic strings contribute to structure formation. Here we apply
Minkowski functionals to 21cm maps which arise if structure is seeded by a 
scaling distribution of cosmic strings embeddded in background fluctuations, and then 
test for the statistical significance of the cosmic string signals using the Fisher combined 
probability test. We find that this method allows for detection of cosmic strings with 
$G \mu >  5 \times 10^{-8}$, which would be improvement over current limits by a factor
of about $3$. 

\end{abstract}

\maketitle 
  
\newcommand{\eq}[2]{\begin{equation}\label{#1}{#2}\end{equation}} 
 
\section{Introduction} 
\label{sec1}

The interest in searching for the signatures of cosmic strings in cosmological
observations has been increasing. This is due in part to the realization that 
cosmic strings can arise in a variety of cosmological contexts. For
example, in many inflationary models (both supergravity-based
\cite{Rachel} and string-based \cite{Tye}) the period of inflation ends with the formation of a 
network of cosmic strings.
Such strings may also be produced in conventional phase transitions which occur after
inflationary reheating \footnote{There are also non-inflationary cosmological models like string
gas cosmology \cite{SGC} in which a network of cosmic strings may be produced.}.
If they are topologically stable, the network of cosmic strings will
persist at all times and will approach a `scaling solution' characterized by a string
distribution which is statistically independent of time if lengths are scaled to the Hubble
radius (see e.g. \cite{ShelVil,HK,RHBrev} for reviews of cosmic strings and their
consequences for cosmology).

Cosmic strings are linear topological defects which arise in a wide range of quantum field
theories as a consequence of a symmetry breaking phase transition. They are lines of
trapped energy density and tension which is equal in magnitude to the energy density.
They are analogous to defect lines in crystals or to vortex lines in superconductors and
superfluids, except for the fact that the cosmic strings arise in relativistic field theories and
hence obey relativistic dynamical equations of motion as opposed to the strings in
condensed matter systems whose dynamics is typically friction-dominated. The trapped energy 
density associated with the strings leads to consequences in cosmology which result in clear 
observational signatures. Since the energy density per unit string length increases
as $\eta^2$ as the symmetry breaking scale $\eta$ increases, cosmology provides the ideal
venue to search for strings at very high energy scales. Thus, searching for strings
in cosmological observations provides an avenue complementary to accelerator experiments
for looking for signals of physics beyond the Standard Model: accelerator signals are
more easily seen for low symmetry breaking scales, while signals in cosmology are more easily detected
for larger values of $\eta$.

Because of the scaling solution of the cosmic string network, cosmic strings lead to a
scale-invariant spectrum of cosmological perturbations, as in inflationary models (see
e.g. \cite{CSearly}). However, the string-induced density fluctuations are highly
non-Gaussian, an effect which is eliminated when computing the usual
power spectrum of density fluctuations. It follows that string signals are much easier to detect in position space than in momentum space. 

The scaling network contains two types of strings, a network of `infinite' strings
\footnote{The network of `infinite' strings can be viewed as a random walk of strings
with a step length comparable to the Hubble scale.} and a distribution of string
loops with radii smaller than the Hubble radius. String loops oscillate because of
their relativistic tension and slowly decay by emitting gravitational radiation. Their
gravitational effects are similar to those of a point mass \cite{CSearly}. In this paper
we are interested in the characteristic non-Gaussian effects of long string segments
and will not further consider cosmic string loops (which produce 21cm signals which
are harder to differentiate from noise \cite{Pagano}).

Due to the relativistic tension of the strings, an infinite string segment will typically have
a velocity in the plane perpendicular to the string which is of the order of the speed of
light \footnote{See \cite{CSnum} for simulations of the dynamics of cosmic string
networks.}. Since space perpendicular to a long straight string is conical with a `deficit angle'
given by \cite{deficit}
\be \label{eq:deficit}
\alpha \, = \, 8 \pi G \mu \, ,
\ee
where $G$ is Newton's constant and $\mu$ is the mass per unit length of the string
(which is proportional to $\eta^2$) , a
string moving with velocity $v_s$ through the matter gas of the early universe at some time $t_i$
will lead to a `wake' \cite{wake}, a thin wedge behind the string with twice the 
background density whose planar dimensions are 
\be \label{eq:planar_dim}
c_1 t_i \times v_s \gamma_s t_i \, .
\ee
Here, $\gamma_s$ is the relativistic $\gamma$ factor associated with the velocity $v_s$,
and $c_1$ is a constant of order one which is proportional to the curvature radius of the
long string network in units of the Hubble radius. The mean thickness $h$ of the wake is
initially
\be \label{eq:in_height}
h \, = \, 4 \pi G \mu v_s \gamma_s t_i
\ee
and for $t > t_i$ it increases by gravitationally accreting matter from above and below the
wake \cite{wakegrowth}.

Current observations (in particular the oscillatory features in the angular power spectrum
of cosmic microwave background (CMB) anisotropies) set an upper bound on the string
tension of the order $G \mu < 1.7 \times 10^{-7}$ \cite{Dvorkin,CSbound}, which corresponds to cosmic strings
contributing less than $5\%$ to the total spectrum of inhomogeneities. We will discuss 
this in more detail in section \ref{sec:obsLimits}. In this paper, we are therefore considering a setup in which there is a contribution of cosmic
strings to the power spectrum in addition to the dominant component of Gaussian,
nearly scale-invariant fluctuations (such as can be produced in inflationary cosmology
\cite{Guth,Mukh} or in string gas cosmology \cite{SGC}).

Cosmic string wakes give rise to distinctive signatures in the large-scale distribution of
matter in the universe. As discussed in \cite{LSS}, they lead to planar structures in the
distribution of galaxies. Since wakes present between the time of last scattering of the
CMB and today represent overdense regions of electrons,
they lead to extra scattering of CMB photons and hence to distinctive rectangular
regions in the sky with extra CMB polarization \cite{Danos1} (with statistically equal
B-mode and E-mode components). The effect relevant to the current paper is that
wakes represent overdense regions of neutral hydrogen and hence \cite{Danos2} 
lead to wedge-like regions in 21cm redshift surveys of extra 21cm absorption or emission
\footnote{Whether the signal is in emission or absorption depends on the ratio of the
gas temperature in the wake and the CMB background temperature.}.  The
amplitude of the position space signal is independent of the string tension $\mu$ and can be
as large as $100 {\rm mK}$ \footnote{The Fourier space
amplitude, on the other hand, does depend on $\mu$ \cite{Wangyi}.} . The width of the wedge, 
however, depends linearly 
on $\mu$.  A crucial point is that wakes exist as non-linear structures at high 
redshifts since as soon as the string passes by, a wake with overdensity 2 forms.
This is to be contrasted to the situation in Gaussian models with a scale-invariant
spectrum of primordial cosmological perturbations in which no nonlinear structures
exist until quite low redshifts. In particular, at redshifts larger than that of reionization, 
the cosmic string signal should stand out against the effects of Gaussian fluctuations
and noise. Thus, 21cm redshift surveys appear to be an ideal window to search
for cosmic string signals.

If cosmic strings exist, then the induced 21cm maps will not only contain the characteristic
wedges of extra absorption or emission. They will also contain noise, most importantly
Gaussian noise from the primordial Gaussian density fluctuations which must be present
in addition to the string-induced perturbations. Good statistical tools will be required in
order to extract the string signals in a quantitative and reliable way.

In this report we study the possibility that Minkowski Functionals \cite{Hadwiger}, 
a tool for characterizing structure which is orthogonal to the usual way of characterizing 
maps using correlation functions, can be applied to identify cosmic string signals in
21cm redshift surveys at reshifts larger than that of reionization. 
The theory behind Minkowski Functionals has its roots 
in Integral Geometry and Hadwiger's Theorem, which states that a $d$-dimensional set of 
convex bodies can be completely described by $d+1$ functionals. Minkowski functionals have been
applied numerous times in cosmology, e.g. to the distribution of galaxies on large
scales \cite{superclusters} and the analysis of CMB maps \cite{CMB}. There have also
been attempts to apply Minkowski functionals to search for signatures of cosmic
strings in the large-scale structure of the distribution of galaxies \cite{CSMink}. 
Since for 21cm redshift surveys the contributions to the `noise' from the
Gaussian source of primordial fluctuations is in the linear regime at the high redshifts
we are interested in (larger than the redshift of reionization), we expect that Minkowski
functionals will be very powerful in extracting cosmic string signals in the case of
21cm maps.

The outline of this paper is as follows: in the following section we outline current observational 
limits on cosmic strings as well as prospects for future detection. In Section III we briefly review 
the toy model for the cosmic string scaling distribution which we use and describe the theoretical 
21cm maps induced by cosmic strings. Section IV then repeats the methodology of section III but 
including a simulation of the background noise. In Section V we review the theory and 
calculation of Minkowski functionals, before presenting our results in Section VI. We conclude with 
a summary of our results and a discussion of future work.

\section{Observational Constraints}
\label{sec:obsLimits}

The strongest constraints on cosmic strings currently come from measurements of the angular
power spectrum of the cosmic microwave background (CMB). The WMAP data provided no 
hints of cosmic strings, and hence placed an upper bound on the string tension
\cite{CSbound}. Specifically, it constrained the string contribution to the
primordial power spectrum to be less than 10\%. However, there are now CMB experiments
with better angular resolution than WMAP provided  and which can yield
improved bounds on the cosmic string tension. The Atacama Cosmology Telescope (ACT) 
\cite{ACT}, a six-meter off-axis telescope with arcminute-scale resolution located in the 
Atacama desert in northern Chile, and the South Pole Telescope (SPT) \cite{SPT} , a 
10 meter telescope located in Antarctica which can study the small-scale angular power 
spectrum of the CMB, both are providing excellent data. A recent analysis of the angular
power spectrum of CMB anisotropies from joint SPT 
and WMAP7 \cite{Dvorkin} made use of a Markov Chain Monte Carlo likelihood 
analysis to place a limit on the string tension of $G\mu < 1.7 \times 10^{-7}$ (at 95\% confidence). 
This analysis was done in the context of the zero width cosmic string toy model which
we will also use in the following. A similar limit was found by \cite{CMBandACT}, who used 
combined WMAP7, QUAD and ACT data to place limits on the tension of Abelian Higgs model
strings. The resulting bound on the string tension was  $G\mu < 4.2 \times 10^{-7}$.

The difference in the results is mostly due to the uncertainties in the distribution of
strings. Due to the very large range of scales involved (the width of a cosmic string
is microscopic but the length is cosmological), numerical simulations of both field
theory and Nambu-Goto strings involve ad hoc assumptions and/or  extrapolations. The
uncertainties in the resulting string distributions appear in free parameters (e.g.
the number of long string segments per Hubble volume and the value of the constant
$c_1$) of toy models distributions of cosmic strings. These uncertainties effect
the angular power spectrum of CMB anisotropies.
 
Local position space searches for signals of individual cosmic strings are 
less sensitive to the uncertainties in the string distribution than power 
spectra. This is the idea behind the proposal of \cite{Amsel, Stewart, Danos0} to search
in position space for the line discontinuities in CMB anisotropy maps 
which  long straight cosmic strings produce due to the Kaiser-Stebbins \cite{KS} 
effect. It was proposed to analyze position space anisotropy maps using of the 
Canny edge detection algorithm. It was shown \cite{Danos0} that
strings with $G \mu \geq 10^{-8}$ might be detectable with this method. Application of this method to the SPT or ACT data may yield interesting results \footnote{See also \cite{Ravi} for 
a different very recent suggestion to search for the string-induced line
discontinuities.}

Another avenue for detection of cosmic strings is via gravitational waves, which will
be produced by oscillating cosmic strings and/or the decay of cosmic string loops. 
Cosmic strings in fact \cite{CSGR} produce a scale-invariant spectrum of gravitational waves
with an amplitude which is similar to or larger than the amplitude of gravitational waves
from simple inflationary models. Pulsar timing \cite{pulsar} or direct detection (e.g. making use of
the Laser Interferometer Gravitational-Wave Observatory (LIGO) \cite{Abbott:2007kv} ) 
provide means for searching for the cosmic string signal. However, at the moment
the bounds are not competitive with bounds obtained from the CMB \footnote{The
gravitational wave signals from cosmic string cusps might be larger, but their
amplitude is very uncertain.}.

We are currently entering a revolution in radio astronomy, with both the 
Square Kilometer Array (SKA) \cite{SKA} and the European Extremely Large 
Telescope (E-ELT)\cite{E-ELT} aiming to be operational by early 2020. Of particular 
interest (in terms of potential to observe cosmic strings) is the SKA, which as the name 
implies will consist of 1 million square meters of collecting area. There are many ongoing 
projects to develop the science and technology necessary to operate the SKA. These are 
categorized into: (1) Precursor Facilities, which will physically be at the SKA site carrying 
out SKA-related projects, (2) Pathfinder Experiments, which will develop SKA-related 
science and technology off-site, and (3) Design Studies, which will investigate technologies 
and develop prototypes. Many of these projects are world class facilities in there own right, 
and we will mention some of the experiments which may have a chance of observing cosmic strings.

One such project is LOFAR \cite{lofar}, a Pathfinder experiment, which is measuring the 
neutral hydrogen fraction of the inter-galactic medium. This will map out the epoch of
reionization between redshifts 11.5 and 6.7, and will be capable out of arc-minute resolution.
For 21cm measurements, the redshift range of LOFAR extends to the ``Dark Ages" before
reionization, redshifts where the cosmic string signal will be cleanest. 
Another SKA project with potential to observe cosmic strings is the MEERKAT experiment 
\cite{Holwerda:2011kd} , which is carrying out a project titled MESMER that will track the neutral 
hydrogen content of the early universe by using carbon monoxide as a proxy. To make an 
accurate prediction as to the clumping of carbon monoxide around a cosmic string would require 
semi-analytical hydrodynamics calculations, but to a rough approximation we should expect to see 
the characteristic wedge of emission/absorption.

\section{The Cosmic String Signal}

\subsection{Modelling the signal}

In any theory which leads to the formation of stable cosmic strings in the early universe,
a network of strings will persist at all later times, and this network will approach a 
`scaling solution', meaning that the statistical properties of the network of cosmic strings 
are independent of time if all distances are scaled to the Hubble radius. The existence of
the scaling solution can be argued for using analytical arguments 
(see e.g. \cite{ShelVil,HK,RHBrev}), and it has been confirmed using extensive numerical
simulations \cite{CSnum}. 

The scaling distribution of strings consists of a network of long strings
of mean curvature radius $\zeta = c_1 t$ (where $c_1$ is a constant of order 1), and 
string loops with a much smaller radius that result from intersection and hence `cutting' 
of the long strings. As a consequence of their relativistic tension, the long strings
have typical translational velocities $v_s$ close to the speed of light. Hence,
Hubble length string segments will intersect with other such segments with
probability close to one on a Hubble time scale. By this process, string
loops are formed and the correlation length of the long string network
(the mean curvature radius) increases in comoving coordianates to keep up
with the Hubble radius, as confirmed in the numerical string evolution
simulations \cite{CSnum}.

As is commonly done in studies of the cosmological consequences of cosmic strings, 
we use an analytical toy model of the infinite string network which was first
introduced in \cite{Periv}: we divide the time interval of interest (usually the
time between the time $t_{eq}$ of equal matter and radiation and the present
time $t_0$) into Hubble time steps. In each time step we lay down an integer
number $N_H$ per Hubble volume of straight string segments of length $c_1 t$
with random midpoints and random tangent vectors. The string segments in
neighboring Hubble time steps are taken to be independent. The long strings
are taken to have a velocity $v_s$ orthogonal to their tangent vectors.
Each such string will produce a wake whose thickness grows in time.

We will be interested in signals which are emitted at a time  
$t_e > t_i$ from string wakes laid down at any time 
$t_i$ between $t_{eq}$ \footnote{Wakes formed at earlier times have smaller
angular extent and thickness. The thickness cannot grow until $t_{eq}$. Hence,
such wakes will be less visible than the ones we study.}
and the present time  . Since the planar dimensions of the wake (whose initial physical
dimensions are given in (\ref{eq:planar_dim})) are constant in comoving
coordinates, their physical size grows with the scale factor. The comoving
thickness of the wake, whose initial physical length is given by
(\ref{eq:in_height}), grows linearly with the scale factor because
of gravitational accretion. Hence, the dimensions of the wake at $t_e$ 
in physical coordinates are given by:
\begin{equation}
\label{dim}
\frac{z_i+1}{z_e+1}
\bigl[ c_1 t_i \times t_i v_s \gamma_s \times 4\pi G\mu t_i v_s \gamma_s\frac{z_i+1}{z_e+1} \bigr]
\, ,
\end{equation}
where $z_e$ is the emission redshift $z(t_e)$.

The dominant form of the baryonic matter in the universe before reionization is
neutral hydrogen (H), which we detect via the 21cm hyperfine line. Measuring
the intensity of the redshifted 21cm radiation from the sky has the potential
of giving us three dimensional maps of the distribution of neutral hydrogen
in the universe (see e.g. \cite{Furl} for an in-depth review article on
21cm cosmology).  As explained in detail in \cite{Danos2}, since wakes are overdense 
regions of neutral hydrogen, they will lead to excess 21cm emission or absorption.
We will see this excess in directions for which our past light cone intersects a wake
at some time $t_e > t_{rec}$, $t_{rec}$ being the time of recombination.
This 21cm signal has the special geometry given by the wake geometry -
a thin wedge in the three-dimensional 21cm redshift survey (see Figure 1), and it is
this special geometry which provides the `smoking gun' signal for cosmic
strings, the signal which we are trying to extract using Minkowski functionals
in this report.

 
\begin{center}
\begin{figure*}
\begin{center}
\label{fig:sketch}
  \caption{The right panel depicts a sketch of the string wake-induced wedge
  of extra emission/absorption in 21cm redshift surveys. The horizontal axis
  represents the two angular directions, the vertical axis redshift. The left
  panel is a space-time sketch showing the position of the string which gives
  rise to the wedge of the right panel. Here, the vertical axis is time, and the
  horizontal axis corresponds to comoving spatial coordinates. Note, in particular,
  the positions where the string wedge intersects the past light cone of an
  observer at the current time $t_0$. The almost horizontal lines represent
  the string at the time when it forms the wake ($t_i$), and the wedge at the
  times $s_1$ and $s_2$ when the back (and front) of the wake intersect
  our past light cone.} 
  \centering
    \includegraphics[scale=0.9]{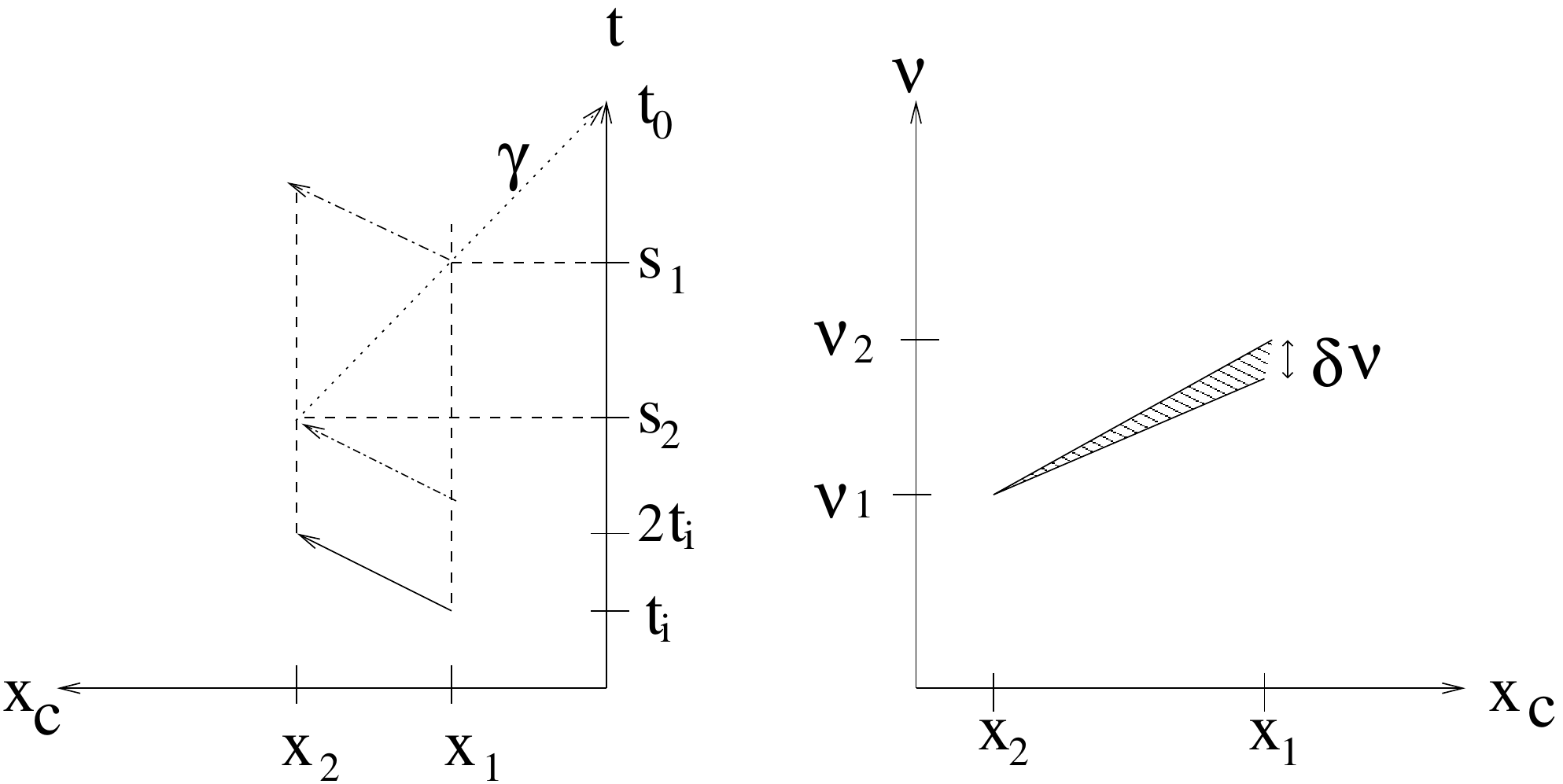}
\end{center}
\end{figure*}
\end{center}


Since already at the initial time $t_i$ the wake is a non-linear density fluctuation, the matter
being accreted onto the wake will collapse onto the wake, shock heat and
thermalize \footnote{This is true if the kinetic temperature which the hydrogen atoms
collapsing onto the wakes acquire is larger than $2.5 T_g$, where $T_g$ is the
temperature of the gas outside of a wake \cite{Danos2}. This will be true for sufficiently
large values of $G \mu$. For smaller values, the wake is diffuse \cite{diffuse}. The density
contrast inside the wake will be smaller, but the thickness will be larger. These effects
will compensate eachother in a Minkowski functional analysis of the wake distribution
provided that the smoothing length in the Minkowski code is sufficiently large.}. 
The hydrogen (H) atoms inside a wake laid down at redshift $z_i$
will have a temperature at time $t_e$ given by:\cite{Danos2}
\begin{equation}
T_K(t_e) \, = \, \frac{16 \pi^2}{75} (G \mu)^2 (v_s \gamma_s)^2 \frac{z_i+1}{z_e+1} \frac{m}{k_B}
\, ,
\end{equation}
where $m$ is the proton mass. Inserting numerical values, and
expressing $G \mu$  in units of $10^{-6}$ and so written as $(G\mu)_6$, we obtain
\begin{equation} \label{eq:TK}
T_K(t_e) \, \approx \, [20 K] (G \mu)^2_6 (v_s \gamma_s)^2 \frac{z_i+1}{z_e+1} \, .
\end{equation}
Note that the numerical simulations by \cite{sornborger} show that the temperature can 
be taken to be approximately uniform inside of the wake. 

The quantity of interest is the 
brightness temperature $T_b$ due to 21cm transitions of the Hydrogen in the wake, or more 
specifically the difference $\delta T_b$ in brightness temperature between photons from 
the wake and those from the surrounding space, 
From this point onwards, `brightness temperature' will refer to $\delta T_b$, defined by
\begin{equation}
\delta T_b(z_e) \, = \, \frac{ T_b (z_e) - T_\gamma (z_e)}{1+z_e} \, ,
\end{equation}
where $T_\gamma (z_e)$ is the redshifted CMB temperature, and $T_b(z_e)$ is the brightness 
temperature due to 21 cm emission in the wake. 
A full derivation which can be found in \cite{Danos2} yields the following
expression for $\delta T_b$:
\begin{equation}
\label{tb}
\delta T_b(z_e) \, = \, 
[0.07 \textit{K}] \frac{x_c}{x_c+1}\left( 1-\frac{T_\gamma}{T_K}\right) \sqrt{1+z_e} \, ,
\end{equation}
where several constants have been absorbed into the prefactor, which can be found in the 
original derivation by \cite{Danos2}. Note that the collision coefficient $x_c$ is given by
\cite{Furl}:
\begin{equation}
x_c \, = \, \frac{n \kappa_{10}^{HH}}{A_{10}} \frac{T_\star}{T_\gamma} \, ,
\end{equation}
where $T_\star$, taken to be 0.068 K, is the temperature corresponding to the 
hydrogen energy splitting $E_{10}$, $A_{10}$ is the spontaneous emission
coefficient of the 21cm transition, $n$ is the number density of hydrogen
atoms, and  $\kappa_{10}^{HH}$ is the de-excitation 
cross section which is given in \cite{Furl}.
Using equation (\ref{tb}), the brightness temperature for a given wake with a given value of 
$z_e$ can be obtained, which in conjunction with the spatial dimensions of the wake, 
allows a 3-D map of brightness temperature to be generated.

\subsection{Generating Temperature Maps}

In the following we will outline the steps in the construction of 21cm redshift maps
in the case of primordial perturbations produced exclusively by strings.
We will consider hypothetical sky maps covering a patch of the sky of angular 
scale $\theta \times \theta$. First, we must compute the number of strings which contribute to
the sky signal in this patch. This is the number of strings whose wakes at some
point in time between $t_{rec}$ and $t_0$ intersect the observer's past light cone
with opening angle $\theta$.

\subsubsection{Distribution of Strings}

According to our model of the distribution of a
scaling string network we divide the time between $t_{eq}$ and the present
time $t_0$ into Hubble time steps. For each such time step centered at
time $t_i$ we now compute the number $N(t_i)$ of string wakes which at some point
in the future of $t_i$ will intersect the past light cone corresponding to an
observational angle $\theta$ (a box of size $\theta \times \theta$ on the sky). 

We first note that $N(t_i)$ is given by the ratio of the comoving volume of the
past light cone 
\be
V_{p}(\theta) \, \simeq \, 27 \bigl( \frac{\theta}{2 \pi} \bigr)^2 t_o^3
\ee
divided by the comoving Hubble volume at time $t_i$
\be
V_H(t_i) \, = \, t_i t_0^2 \, ,
\ee
multiplied by the number $N_H$ of strings per Hubble volume in the scaling
solution (a number which according to numerical simulations is in the range
$1 < N_H < 10$ \cite{CSnum}. This gives
\be \label{stringcount}
N_i \, \simeq \, N_H 27  \bigl( \frac{\theta}{2 \pi} \bigr)^2 z(t_i)^{3/2} \, .
\ee

Now that we know the total number of string wakes produced at redshift $z_i$
which will be seen in the simulation box, we need to determine the
distribution of the redshift $z_e$ at which they will cross the past
light cone of the simulation region in the sky. This can be
determined by remembering that light travels in a straight line in
conformal coordinates (comoving spatial coordinates and conformal
time $\tau$). Hence, the number of wakes which intersect the past
light cone at conformal time $\tau$ (where the conformal time is
measured backwards, meaning $\tau = 0$ at the present time
and $\tau$ increases as we go back in time) in a conformal time interval
$d \tau$ is
\be
d N_i(\tau) \, = \, \bigl( \frac{\tau}{\tau_{eq}} \bigr)^2 {\cal N} d \tau \, ,
\ee
where ${\cal N}$ is determined by demanding that the integral over
$\tau$ from $\tau_i \equiv \tau(t_i)$ to $\tau(t_0)$ gives $N_i$.
The result can be integrated from $0$ to $\tau_e$ to yield the
number of string wakes $N_i(\tau_e)$ which intersect the past light
cone at a time later or equal to $\tau_e$:
\be \label{stringcount2}
N_i(\tau_e) \, = \, \left( \frac{\tau_e}{\tau_i} \right)^3 N_i \, .
\ee
To determine the redshift distribution of emission times $z_e$, we
need to relate the conformal time $\tau_e$ to the emission redshift $z_e$. 
We start by expressing $\tau_e$ in terms of $t_e$,
\be 
\mathrm{d}\tau_e = \frac{\mathrm{d}t_e}{a(t_e)} \, ,
\ee
which evaluates to 
\be 
\label{eq:taueze}
\tau_e = 3t_0 \left[1-\left(z_e+1\right)^{-\frac{1}{2}}\right] \, .
\ee
An expression for the number $N_i (z_e)$ of string wakes with an
emission redshift larger or equal to $z_e$ can now be obtained by 
substituting the preceding equation into equation \ref{stringcount2}, 
and using our expression for $N_i$ from \ref{stringcount}. This gives
\be 
N_i (z_e) \, =  \, N_H 27  
\bigl( \frac{\theta}{2 \pi} \bigr)^2 z(t_i)^{3/2} \left[\frac{1-\left(z_e+1\right)^{-\frac{1}{2}}}{1-\left(z_i+1\right)^{-\frac{1}{2}}}\right]^3 
\ee

Due to the discrete nature of this simulation, we need to approximate 
this relationship by splitting the redshift space into slices, and laying 
down strings in each slice. The smallest possible slice would be the number 
of discrete values of redshift in the simulation (128, as we will discuss 
shortly) divided by the range of redshift, which we shall take to be 50 for this 
example. However, as we must have an integer numbers of wakes, a slice this 
small would result in the majority of slices having 0 wakes. Hence we 
instead choose a slice thickness $\Delta z_e = 1$, which will contain a 
number of wakes given by
\bea \label{stringdist}
& & \Delta N_i(z_e, z_e-1) \, = \, 
\left[ 1.41\times10^6 \right] \\
& & \,\,\,\,\,\, \left[ -2(x+1)^{-1/2}+2x^{-1}-\frac{2}{3}x^{-3/2}  \right]_{z_e-1}^{z_e} \, ,
\nonumber
\eea
where we have included all prefactors into the constant. A plot of this 
distribution is shown in Figure 2. As is apparent, the largest number
of wakes intersect the past light cone at low redshifts. This can
be understood since the low redshift range covers most of the
physical volume.

\begin{center}
\begin{figure}
\begin{center}
\label{fig:sketch2}
  \caption{Distribution of the redshifts at which the string wakes
intersect the past light cone (for a fixed time $z_i = 10^3$ at
which the strings were produced). The horizontal axis is the
emission redshift $z_e$, the vertical axis is the number 
$\Delta N_i(z_e, z_e - 1)$ of string
wakes produced at redshift $z_i$ (taken to be $10^3$) which intersect
the past light cone of the sky area we are considering in the
redshift interval between $z_e - 1$ and $z_e$.} 
  \centering
    \includegraphics[scale=0.4]{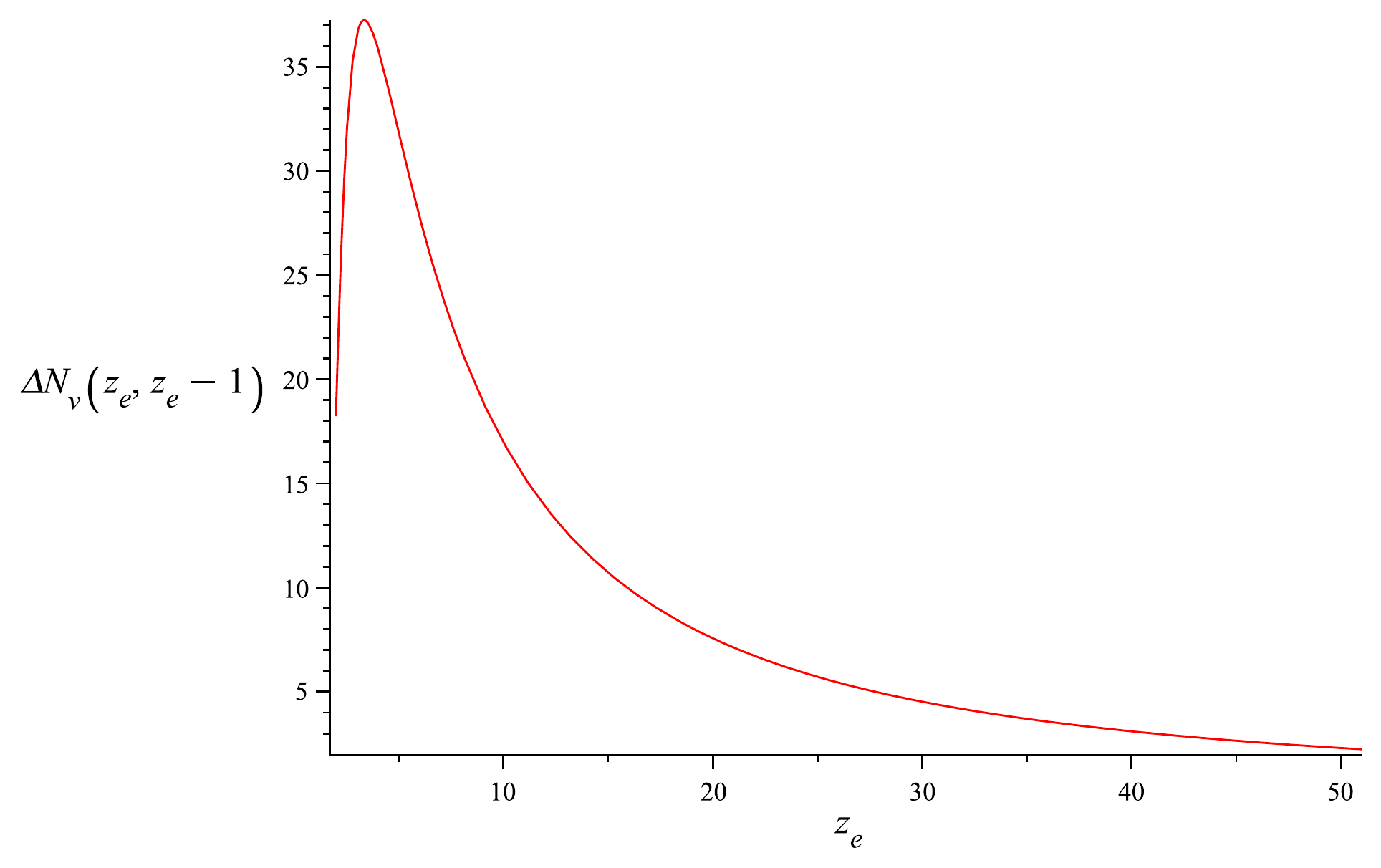}
\end{center}
\end{figure}
\end{center}

\subsubsection{Angular and Redshift Scales of a Wake-Induced Wedge}

As discussed in Section 2, each string wake which intersects the past
light cone of the observer's angular patch will lead to a thin wedge
of extra 21cm absorption or emission.  We are
concerned with string wakes produced at time $t_i$
which intersect the past light cone of the observational patch at some time 
$t_e > t_{rec}$. Each such string wake will lead to a thin wedge in a 
21cm redshift survey of a certain angular extent and certain thickness.

First, let us consider the angular extent of the wedge. A string laid down at a time $t_i$ 
will create a wake with physical dimensions given by equation~\ref{dim}. 
To calculate the angular size of a wake, one must start with the physical length of the 
wake at time $t_i$ which in direction tangent to the string is:
\begin{equation}
x_p(t_i) \, = \, c_1 t_i  \, ,
\end{equation}
where (as we recall from before) $c_1$ is a constant of order 1. In
the direction of string motion the size is the same except that the
factor $c_1$ gets replaced by $v_s \gamma_s$. In our work
we choose the rather realistic parameter values where the two
constants are the same. The corresponding comoving distance is then
given by
\begin{equation}
x_c(t_i) \, = \, \left(\frac{t_0}{t_i}\right)^{\frac{2}{3}} x_p(t_i) = (z_i +1)^{-1/2}c_1 t_0  \, .
\end{equation}
Hence, the corresponding angle $\theta_i$ is:
\begin{equation}
\theta_i \, = \, (z_i +1)^{-1/2}c_1 \times 90^o  \, .
\end{equation}
For the most numerous and thickest wakes, $z_i+1=1000$, 
and if the constant $c_1$ is taken to be $1/3$ , the answer 
is then approximated by:
\begin{equation}
\theta_i \, = \, (z_i +1)^{-1/2} \times 30^o \,   
\end{equation}
which yields approximately $1^o$ as the angle of the wake-induced wedge. 

Since the string which produces the wake is moving, the string-induced wake is
slightly ``tilted" in the three-dimensional redshift-angle space, i.e. the tip of the
wedge (which corresponds to the string at the latest time) is at a slightly larger
redshift than the mean redshift of the tail of the wedge (where the string was at
the initial time). This does not change the fact that the projection of the wedge
onto the angular plane has the dimension given above. However, when
computing the projection into the redshift direction we have to be careful.
The wedge is thin in direction perpendicular to the plane of the wedge, but
this perpendicular direction is at an angle relative to the redshift direction.
In the following we will compute the thickness of the wedge in perpendicular
direction to the wedge plane.

We now calculate the thickness of the wake in the perpendicular direction. 
The finite thickness in redshift direction originates from the fact that photons
originating from different parts of the wake are redshifted by slightly different
amounts. The starting point of the computation is the formula
\begin{equation}
z + 1 \, = \, \left( \frac{t}{t_0} \right)^{\frac{2}{3}}
\end{equation}
valid in the matter-dominated period. Taking differentials, and setting $t = t_e$ yields
\begin{equation}
\label{eq:diff}
\Delta z \,= \, \left( \frac{2}{3} \right) t_0^{-2/3} t_e^{-1/3} \Delta t_e \, , 
\end{equation}
where $\Delta t_e$ is the time delay between photons from the top and the bottom
of the wedge. We compute $\Delta t_e$ at the midpoint of the wake (point of medium
thickness). This is given by the thickness of the wake, which, as discussed
in \cite{Danos2} grows linearly in comoving coordinates because of
gravitational accretion onto the initial overdense region. 

Using linear perturbation theory, the width of the wake at time $t_e$ is
\be
w(t_e) \, = \, 4 \pi G \mu v_s \gamma_s t_0 \bigl( z(t_i) + 1 \bigr)^{1/2} \bigl( z(t_2) + 1 \bigr)^{-2} \, ,
\ee
which equals $\Delta t_e$. Hence, from (\ref{eq:diff}) it follows that
\be
\Delta z \, = \, \frac{8 \pi}{3} G \mu v_s \gamma_s \bigl( z(t_i) + 1 \bigr)^{1/2} \bigl( z(t_2) + 1 \bigr)^{-3/2}
\, .
\ee
The analysis of gravitational accretion performed using the Zel'dovich approximation
instead of with naive linear cosmological perturbation theory yields the same
result \cite{Danos2} except that the coefficient is $24 \pi / 5$ instead of $8 \pi / 3$.

Once the dimensions of the wake are calculated, it remains to map this onto a 3-D lattice 
of points from which the Minkowski Functionals can be calculated. To do this requires setting
the resolution with which the wakes will be studied, and then scaling up the 
dimensions of the wake. For the purposes of this simulator, we use a 
lattice of side length $128$, and study a 
volume that subtends an angle of $10^o$ in both the angular directions, with  $z_e$ ranging 
from 5 to 50, and $N_H = 3$. In this case, the angular resolution is $10^o/128=.078^o$, while the resolution
 in the redshift direction is $45/128=.35$. To scale up the dimensions of the wake only requires 
multiplying each dimension by the inverse of the resolution (ie: $128/45=2.84$ in the $z_e$
direction).

\subsubsection{Brightness Temperature: $\delta T_b$}

The final step in the calculation is to determine the brightness temperature at every 
point of the three dimensional angle-redshift map $(\theta, \phi, z)$. Any point for 
which the past light ray in angular direction $(\theta, \phi)$ does not
intersect a wake at redshift $z$ yields zero brightness temperature. A
point which at redshift $z_e$ is intersecting a wake due to a string
present at time $t_i$ is assigned a
non-vanishing brightness temperature given by Equation (\ref{tb}), with
$T_K$ obtained from Equation (\ref{eq:TK}).  Overlapping wakes are assumed 
to be non-interacting.

\section{Background Noise}

\subsection{Modelling the Signal}

As discussed in the introductory section, in our setup the cosmic strings only make
a contribution of less than $5\%$ to the total power spectrum of inhomogeneities.
The dominant contribution is in the form of an approximately scale-invariant spectrum of
Gaussian perturbations such as can be produced in a number of cosmological
scenarios, e.g. in inflationary cosmology \cite{Mukh} or in string gas cosmology \cite{SGC}.
The dominant Gaussian fluctuations are in the linear regime until late times on large 
length scales and are hence not expected to contribute a lot to 21cm fluctuations at
high redshifts (redshifts before reionization). There will, however, be smaller
scale fluctuations which become non-linear, forming so-called ``mini-halos" which
then contribute to the spectrum of 21cm fluctuations. It is crucial for us to check
that the cosmic string effects can be detected above the noise from the
Gaussian fluctuations (which we call ``background noise" in the following).

The contribution of ``background noise" to the 21cm fluctuations of the universe has been 
studied in detail (see e.g. \cite{shapiroAnalytical, arXiv:1005.2502}). As was shown,
there is an effect on 21cm maps which comes from the diffuse inter-galactic
medium (IGM). However, this effect is homogeneous in space on the cosmological
scales which interest us here and will hence not be further considered in this
paper. Instead, we focus on the contribution of the inhomogeneities mentioned above which
lead to the formation of mini-halos. This contribution has been studied done both 
semi-analytically, \cite{shapiroAnalytical}, and later using large scale numerical simulations 
\cite{arXiv:1005.2502}.

To calculate the signal from minihalos we must find the mass function 
$\mathrm{d}n/\mathrm{d}M$, which can, when integrated over a range of masses, give the 
number density for minihalos in this mass range. There are two possible mass profiles for 
minihalos, the Sheth-Tormen model and the Press-Schechter model. Recent numerical 
simulations support the Press-Schecter model, (see Figure 2 of \cite{arXiv:1005.2502}). 
Once we know the mass function, the brightness temperature is then calculated using 
line of sight integrals given a realization of the inhomogeneities described by the
mass function.

The Press-Schecter function for the number density of minihalos is given 
by \cite{Press:1973iz}:
\begin{equation}
N(M) \mathrm{d}M = -\left( \frac{\bar{\rho}}{M}\right)\left( \frac{2}{\pi}\right)^{1/2} \frac{\delta_c}{\sigma^2}\frac{\mathrm{d}\sigma}{\mathrm{d}M} e^{-\frac{\delta^2_c}{2 \sigma^2}} \mathrm{d}M
\end{equation}
where $\bar{\rho}$ is the mean (baryonic and dark) matter density of the universe, 
and $\sigma^2(M,t)=\langle \delta^2_M \rangle (t)$ is the root mean square
mass fluctuation on the mass scale $M$. Halos will form in regions with a $\delta_M$ above 
a critical fluctuation $\delta_c \approx 1.69$.  Our number density simplifies to:
\begin{eqnarray}
N(M,z)  \mathrm{d}M &=& \Omega_m \left(\frac{(3+n)}{6} \right)\left( \frac{\bar{\rho}}{M^2}  \right) \nonumber \\ 
&& \times\left( \frac{2}{\pi}\right)^{1/2} \left(\frac{M}{M_* (z)}\right)^{\frac{(3+n)}{6}} \\
&& \times \exp\left[{-\frac{1}{2}\left(\frac{M}{M_*(z)}\right)^{(3+n)/3}}\right] \mathrm{d}M \nonumber
\end{eqnarray}
where $M_*$ is the mass scale for which the fractional density fluctuation
equals $\delta_c$, and where we assume a primordial power spectrum 
of density fluctuations $P(k) \approx k^n$ (here, $P(k)$ does not contain the
phase space factor $k^3$ and is not dimensionless. A scale-invariant
spectrum of curvature fluctuations corresponds to $n = 1$). 

The spatially averaged brightness temperature of a set of halos is given 
by \cite{shapiroAnalytical}:
\begin{equation}
\label{eq:Tbz}
\bar{ \delta T_b } = \frac{c(1+z)^4}{\nu_0 H(z')} \displaystyle \int_{M_{min}}^{M_{max}} \Delta \nu_{eff} \delta T_{b,\nu_0} A \frac{\mathrm{d}N}{\mathrm{d}M} \mathrm{d}M
\end{equation}
where $\Delta \nu_{eff}$ is the redshifted effective line width, $\delta T_{b,\nu_0} $ 
is the brightness temperature of individual minihalo at the frequency $\nu_0$, and 
A is the geometric cross section of the minihalo. The integral (worked out in
\cite{Iliev:2002gj} - see Figure 3 in that paper) yields a noise temperature 
which peaks at a value of 4mK at a redshift of 10 and decays at higher
redshifts.

\subsection{Generating Noise Temperature Maps}

Given that the signal from the IGM is spatially uniform, we focus on the contribution from
minihalos. The corresponding noise temperature maps are constructed by
modeling the spatial fluctuations as a three-dimensional Gaussian random field
with a power spectrum $P_T(k)$ (again defined without the phase space factor
$k^3$) with a slope which will be discussed
below, and with a variance given by (\ref{eq:Tbz}). The
amplitude of the resulting noise is rescaled in redshift direction by
incorporating the redshift dependence of the variance from
(\ref{eq:Tbz}). 

To accomplish this, we follow the procedure laid out in \cite{Danos0},
which we generalize to 3 spatial dimensions,  We can expand
the spatial fluctuations in temperature into hyperspherical harmonics.:
\begin{equation}
\frac{\Delta T}{T} (\theta , \phi , \psi) \, = \, \displaystyle \sum_{l , m, n}
a_{l,m,n}  Y_{l,m,n} (\theta, \phi, \psi),
\end{equation}
where the $Y_{l,m,n}$ are the spherical harmonics generalized to 3+1
dimensions, the `\emph{hyperspherical harmonics}', and the $a_{l,m,n}$
are the coefficients of this expansion.
We can simplify this expression to a sum of plane waves, by making the 
flat sky approximation (valid for angular scales $< 60^o$). This gives us:
\begin{equation}
  \tilde{T}(\vec{x}) \, = \, \displaystyle \sum_{\vec{k}} \tilde{T}(\vec{k}) 
  e^{i \vec{k} \cdot \vec{x}}
\end{equation}
where we have used the abbreviation 
$ \tilde{T}(\vec{x}) \equiv \frac{\Delta T}{T} (\vec{x})$. 
Comparing this decomposition with our previous 
expression in terms of hyperspherical harmonics, we see that the $\tilde{T}(\vec{k})$ 
correspond to the $a_{l,m,n}$ and hence the $\tilde{T}(\vec{k})$ give us the 
power spectrum:
\begin{equation}
\langle \tilde{T}(\vec{k})^2 \rangle \, = \, \langle a_{l,m,n} ^2 \rangle \, \sim \, P_T(k)
\end{equation}
where the temperature power spectrum $P_T$ is defined by the last equality.

Since the mini-halos are produced by the density fluctuations we use the
power spectrum of density fluctuations to determine the slope of $P_T(k)$.
Thus, we take $P_T(k)$ to be proportional to the primordial power
spectrum:
\be
P_T(k) \, \propto \, k^n \,\,\, {\rm with} \,\,\, n = 1 \, .
\ee
 and the amplitude is determined by demanding that the variance is given
 by (\ref{eq:Tbz}). First, in fact, we generate a three-dimensional
 Gaussian random field $T_{GRF}$ with variance 1.

Note that in a model with a primordial scale-invariant spectrum of curvature
perturbations, the dimensionless power spectrum of fractional density
perturbations is approximately scale-invariant on scales which entered
the Hubble radius before the time of equal matter and radiation. This is
a consequence of the processing of the primordial spectrum which
happens on sub-Hubble scales (see e.g. any text which discusses
the Newtonian theory of cosmological perturbations, e.g. \cite{Peebles}).
The spectrum on small scales thus takes on a slope $n = - 3$. We do
not consider this effect in the current simulations.

Using the ergodic hypothesis, we construct a Gaussian noise field by
drawing the $\tilde{T}(\vec{k})$ for fixed $k = | \vec{k}|$ from a Gaussian
distribution with power given by $P_T(k)$ and amplitude chosen
to give variance 1. 

We do our calculations on a lattice of $(\tilde{k}_1, \tilde{k}_2, \tilde{k}_3)$ 
coordinate values ranging from $0$ to $N_{max}-1$. 
We convert to the corresponding k values with:
\begin{equation}
k_i \, = \, \frac{2 \pi}{L} (\tilde{k}_i - k_{max})
\end{equation}
where $k_{max}$ corresponds to the angular resolution of the lattice in the
$k_i$ direction.

We can then calculate the $\tilde{T}(\vec{k})$ using:
\begin{equation}
\tilde{T}(\vec{k}) \, = \, \sqrt{\frac{P_T(k)}{2} }\left(g_1(\vec{k}) + ig_2 (\vec{k}) 
\right)
\end{equation}
where $g_1$ and $g_2$ are randomly generated from a Gaussian distribution 
with variance 1 and mean 0, and we enforce $\tilde{T}(\vec{k})=\tilde{T}(-\vec{k})$ 
to ensure that $T(\vec{x})$ is real. 

We can then construct a spatial map by taking the inverse Fourier 
transform of the preceeding expression. The result of this is a Gaussian
random field $T_{GRF}$ with a correlation function specified by the power
spectrum and variance 1. We now wish to enforce the redshift dependence of the amplitude,
which we do by identifying redshift $z$ with the third spatial direction.
Hence the final noise field is specified by:
\begin{equation}
T_{noise} (x,y,z) = \bar{\delta T_b}(z) T_{GRF} (x,y,z)
\end{equation}
with  amplitude $\bar{\delta T_b}(z) $ given by (\ref{eq:Tbz}). Note that since
the change in amplitude is uniform over the angular directions, this introduces
deviations from strict Gaussianity of the the three-dimensional distribution.

\section{Integral Geometry and Minkowski Functionals}

Minkowski Functionals are a useful tool to analyze the topology of a 
$d$-dimensional map. Originally, these functionals were considered
in the context of describing the topology of a body $B$ embedded in a $d$-dimensional Euclidean
space using an approximation in which the body $B$ was approximated by
the set of convex bodies of a particular form but varying size (hence we
have functionals and not just functions).  In our application we will
consider the Minkowski functionals as functionals of the class of bodies (termed an `excursion set')
which enclose regions of the map where the value of the variable which is mapped 
is larger than a cutoff value and we consider varying the cutoff value.
For example, applied to CMB temperature anisotropy maps we consider
bodies which are regions of the sky where the temperature $T$ is larger
than a cutoff temperature $T_c$ whose amplitude we vary. In the case of
interest here we consider three dimensional 21cm brightness temperature
maps and we consider the topology of the volumes which contain points
where the temperature exceeds a critical temperature whose magnitude we
vary.
 
Minkowski first developed these functionals \cite{Minkowski}
in 1903 to solve problems of  stochastic geometry \cite{Schmalzing1}. This
then led to the 
development of Integral Geometry in the mid 1900's. At the heart of Integral Geometry 
is Hadwiger's Theorem \cite{Hadwiger}, which deals with the problem of
characterizing the topology of the body $B$ using scalar 
functionals $V$. These functionals must satisfy certain requirements \cite{Schmalzing2}:

\begin{enumerate}
\item {\bf Motion Invariance} : The functionals should be independent of the position 
and orientation in space of the body.
\item {\bf Additivity}: The functionals applied to the union of two bodies equal the
sum of their functionals minus the functionals of their intersection:
\begin{equation}
	V(A\cup B) \, = \, V(A) + V(B) - V(A\cap B) \, .
\end{equation}
\item {\bf Conditional Continuity}: The functionals of convex approximations to a convex 
body converge to the functionals of the body.
\end{enumerate}

Hadwiger's Theorem \cite{Hadwiger} states that for any $d$ dimensional convex 
body, there exist $d+1$ functionals that satisfy these requirements, denoted 
$V_j \, , j = 0, ... d$, for the $j$'th functional. Furthermore, these functionals provide a 
\emph{complete} description of topology of the body.  Mathematically, the $j$'th functional of a 
$d$-dimensional body $B$ is an integral over a $(d-j)$-dimensional surface of $B$. 
For example, in three dimensions, the first functional $V_0$ is simply the volume of the 
body, and the second ($V_1$) is the surface area. In all dimensions, the last 
functional $V_j$ is given by the Euler characteristic 
$\chi$, which in three dimensions is defined \cite{Schmalzing2} as:
\begin{eqnarray}
\chi =&& \textit{number of components - number of tunnels} \nonumber \\
&& \textit{+ number of cavities} \, .
\end{eqnarray}
The simplest example is a set of balls of radius $r_i$. When $r_i$ is very small, 
there are no intersections of balls, and $\chi$ is very close to the number of balls.  
As $r_i$ increases, the balls will begin to intersect, creating tunnels. Thus $\chi$ 
will become negative. At a certain threshold value of $r_i$, $\chi$ will once again 
be positive, as tunnels are cut off to form cavities. Finally, when $r _i$ becomes
very large, the entire space is filled, and $\chi$ is equal to 0.

\begin{center}
\begin{table}
\label{tab:functionals}
\caption{Geometric Interpretation of Minkowski Functionals in 1,2, and 3 dimensions \cite{Schmalzing1}}
\centering
    \begin{tabular}{c c c c}
    \hline
d                  &   1              & 2 & 3     \\ \hline
$V_0$                  &   length              & area & volume     \\ 
$V_1$                  &    $\chi$            & circumference & surface area     \\ 
$V_2$                  &     -            &  $\chi$ & total mean curvature     \\ 
  $V_3$                  &     -            &  - &   $\chi$   \\  \hline
\end{tabular}
\end{table}
\end{center}

The geometric meaning of the functionals for $d \leq 3$ is summarized in 
Table I \cite{Schmalzing1}.
A key feature of Minkowski Functionals is that they incorporate information about
correlation functions of arbitrary 
order \cite{Schmalzing1}. Hence, the Minkowski Functionals are much more sensitive to 
non-Gaussianity then a three or four point correlation function, making them an ideal tool 
to search for signatures of topological defects such as cosmic strings. 
An important fact is that exact expressions exist for a Gaussian Field (given by Tomita 
in \cite{Tomita}), allowing for a visual comparison between string-based and
Gaussian maps (since all we have to look for is statistically significant deviations
from the curves for a Gaussian distribution). We will apply this method in Section 6. 

Next we will apply Minkowski functionals
to the temperature maps whose construction is described in the previous
section. For any fixed value of the temperature, we consider
the corresponding iso-temperature surface. The Minkowski functionals
probe the topology of this set of surfaces (or equivalently the bodies they enclose).
They are calculated as a function of the temperature threshold, and the results
are plotted as a function of this threshold. It is standard practice to use as the 
x axis not the actual value of the field, but a variable $\nu = f(x)/\sigma$ which is the number of
standard deviations $\sigma$ by which the value of the field $f(x)$ (scaled to have zero spatial average) deviates from zero .
Hence, a rescaled threshold $\nu=\pm 1$ implies a fluctuation of the 
field by one standard deviation from the average value. 
The reason for using $\nu$ is to remove the effect 
of a constant factor on the functionals, and hence to allow for a comparison 
with structure that may be identical in every way but the magnitude
of the temperature. For the purposes of this simulator, we calculated the functionals for 
the range $-4\leq \nu \leq 4$.

The calculation of the Minkowski functionals was done using the 
program \emph{Minkowski3}, written by Thomas Buchert. This piece of software 
takes a 3-D array of floats as input (in binary format), and outputs two 
estimates (with error values) of the Minkowski Functionals as well as the 
functionals for a Gaussian field.  The two estimates for the Minkowski functionals 
correspond to two methods, one derived from differential geometry and the other 
from integral geometry. As a full derivation of each can be found in 
\cite{minkowskicalculation}, only a brief explanation will be provided here.

In the context of differential geometry, it is possible to describe all local curvature 
in terms of geometric invariants (Koenderink Invariants). We then express 
the global Minkowski functionals $V_k$ in terms of the local Minkowski 
functionals $V_K^{(loc)}$ , which are calculated in \cite{minkowskicalculation}. 
For the three dimensional case, the functionals of a field $\nu({\bf x})$ in 
a volume $\mathcal{D}$ are given by
\begin{equation}
V_k(\nu) \, =  \, 
\frac{1}{|\mathcal{D}|} \int_{D} d^3x \delta (\nu - \nu({\bf x})) \; |\nabla \nu({\bf x})| \; V_k^{(loc)}(\nu,{\bf x})
\, .
\end{equation}
In the discrete case, this is equivalent to summing over the lattice and taking a spatial average.

In the context of integral geometry, the Minkowski Functionals are calculated 
using Crofton's Formula. This provides a very elegant expression for the 
$k$'th functional of a $d$ dimensional object K, in a volume that consists of 
$L$ lattice points of cubic lattice spacing $a$:
\begin{equation}
V_k^{(d)}(K) \, = \, 
\frac{\omega_d}{\omega_{d-k} \omega_k} \frac{1}{a^k L}\displaystyle\sum\limits_{j=0}^k (-1)^j \frac{k!(d-k+j)!}{d!j!} N_{d-k+j}(K) \, ,
\end{equation}
where $\omega_j$ is the volume of a $j$-dimensional unit sphere, and 
$N_{j}(K)$ is the number of $j$-dimensional lattice cells contained in K. 
For example, $N_0(K)$ is the number of lattice points in K, and $N_3(K)$ 
gives the number of cubes contained within K.

\section{Results}
\subsection{Functionals of Cosmic Strings}

We first consider the Minkowski Functionals of the wake signal alone, by taking a scaling 
solution of strings with $(G\mu)_6=0.17$, using the notation $(G\mu)_6= G \mu \times10^{-6}$. 
The results are shown in Figure 3. The four boxes show the mean Minkowski functionals 
after 100 simulations, with the errors taken to be the standard deviation. In each box, the green 
curve is based on calculating the functionals using  the method of Koenderink Invariants, and the 
red curve is based on using the Crofton Formula (the red and green curves are almost perfect 
replicas of one another, and so the green is only visible in regions where they differ). 

Since the number density of wakes is peaked around high redshift, the mean brightness 
temperature fluctuation $\delta T_b$ will be negative, and hence the threshold value 
$\nu = 0$ corresponds to a slightly negative brightness temperature fluctuation. The
threshold value $\nu_c$ corresponding to $\delta T_b = 0$ is
a small positive number. It is at this threshold value that the
four Minkowski functionals change abruptly. For $\nu > \nu_c$
the volume with $\delta T_b$ greater than the threshold value
vanishes, and hence the functionals $V_1, V_2$ and $V_3$ vanish.
For $\nu < \nu_c$ the high temperature volume does not vanish,
but it corresponds to the outside of the wakes as opposed to the
inside. Hence, the integrated mean curvature is negative. The key feature to notice in 
these plots is the abrupt change that occurs at the threshold value and the asymmetry 
about this threshold.

\begin{center}
\begin{figure*}
\begin{center}
\label{fig:wakes}
  \caption{Minkowski functionals of a pure scaling solution of cosmic string wakes, with 
  $G\mu=0.17\times 10^{-6}$. Each of the four boxes shows
  the results for one of the four functionals, averaged over 100 simulations each 
  using a lattice of $128^3$ pixels.  Each box contains two curves.
  corresponding to computing the functionals computed making use of 
  one of the two methods, either Koenderink Invariants (the green curve) or the 
  Crofton Formula (the red curve).}
  \centering
    \includegraphics[scale=1.2]{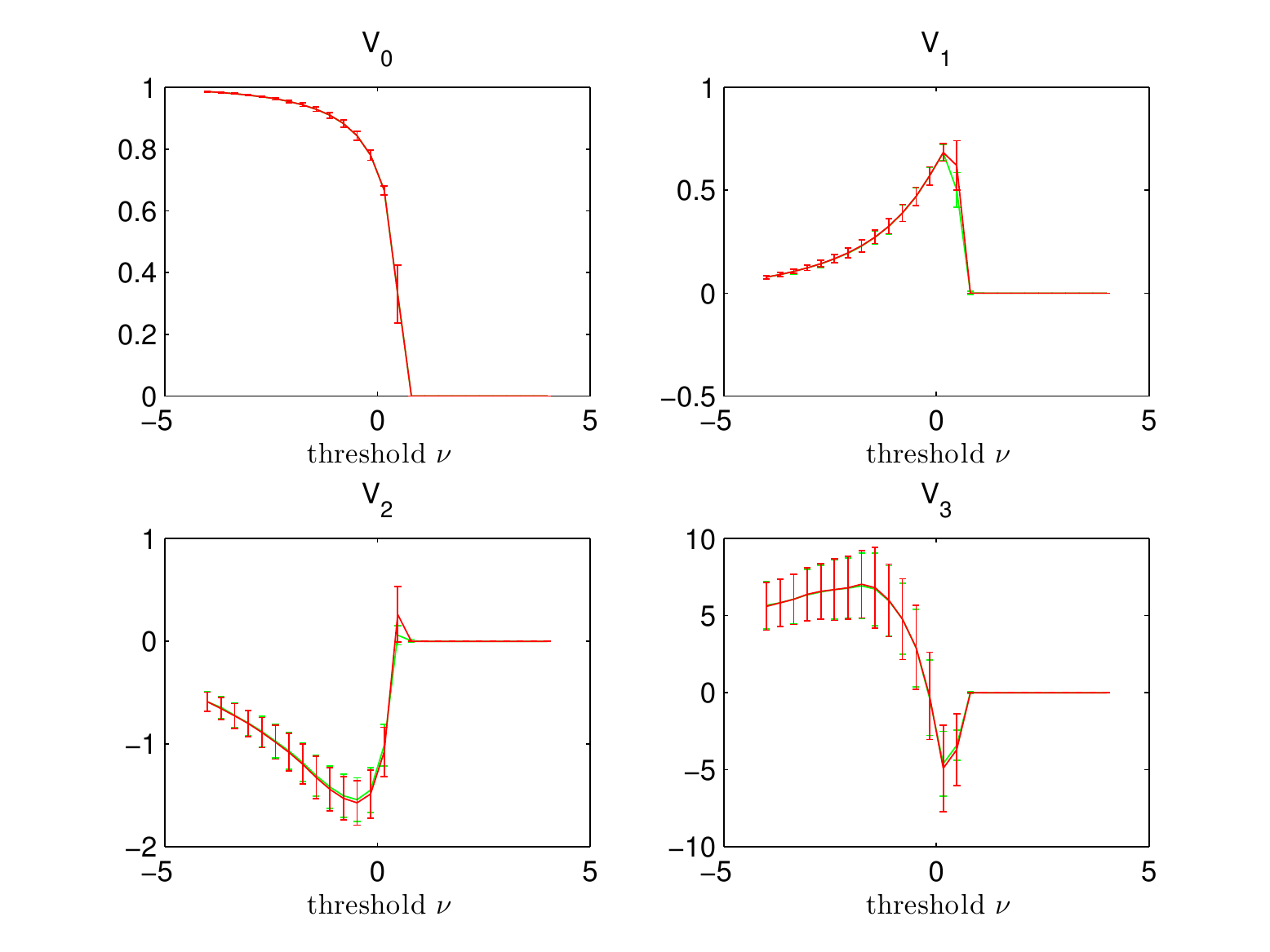}
\end{center}
\end{figure*}
\end{center}

\subsection{Functionals of Noise}

We now turn our attention to the Minkowski Functionals of the background noise due to 
minihalos, shown in Figure 4, where we again calculate the average functionals over 
100 simulations and indicate the standard deviation with error bars. The corresponding 
functionals for a Gaussian random field are shown as the dashed black curves, whereas
the  functionals for the noise maps are shown in red and green. 
The key feature of these plots is the symmetry (or antisymmetry in the case of $V_2$ and 
$V_0$) about 0, which matches the results for a  Gaussian random field. 
However the noise map functionals all share the common characteristic of being tightly 
concentrated about 0, which allows the noise map to be differentiated from a pure Gaussian 
random field. As mentioned above, the deviation of the noise maps from being a pure
Gaussian random field comes from the scaling of the amplitude with redshift which is
uniform over the two angular coordinates.

\begin{center}
\begin{figure*}
\begin{center}
\label{fig:noise}
  \caption{Minkowski functionals of the background noise, averaged over 100 simulations 
  each using a lattice of $128^3$ pixels.  Each box contains three curves.
  Two of them are for the noise map with the functionals computed making use of 
  one of two estimates, either Koenderink Invariants (the green curve) or the 
  Crofton Formula (the red curve). The dashed black curves give the Minkowski
  functionals for a Gaussian map.}
  \centering
    \includegraphics[scale=1.2]{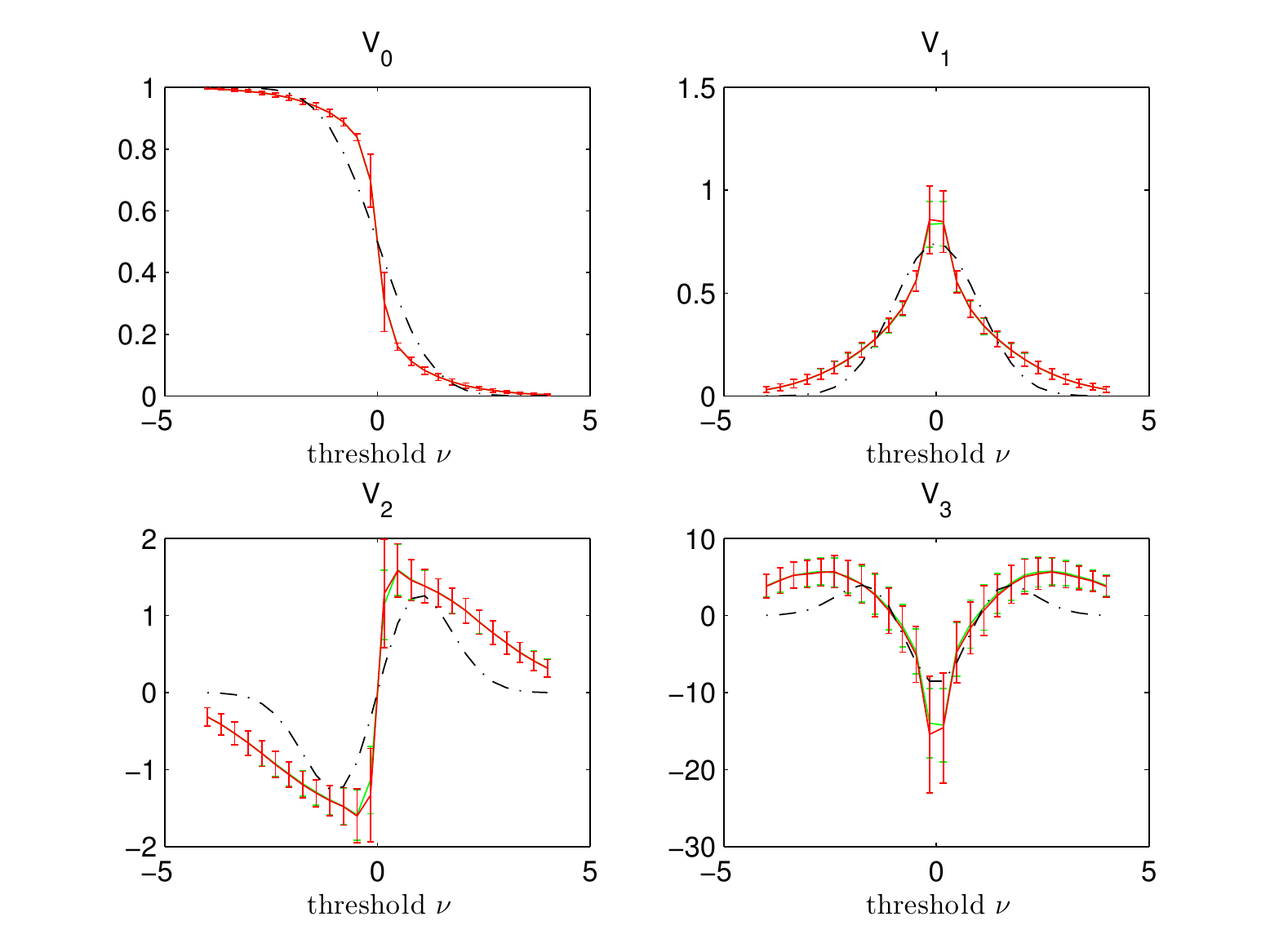}
\end{center}
\end{figure*}
\end{center}

\subsection{ Functionals of Cosmic Strings Embedded in Noise}

Armed with our knowledge of the behaviour of the Minkowski Functionals for both pure cosmic 
string wake and background noise maps, we can start with the real fun: differentiating maps of 
wakes and noise from maps of pure noise. To do this, we generate a map of wakes and 
noise by adding (at every point in space) the temperature fluctuations from wakes and the 
fluctuations from the mini-halo noise maps. We can then compare the Minkowski functionals 
as we vary the string tension, allowing us to place constraints on the minimum string tension 
necessary for the strings to be detected via this method.

For each value of the string tension $G \mu$ we performed 100 simulations.  
For each simulation we computed the 
$\chi^2$ significance value of the difference between the string and a pure noise map. 
This was done in the following way:
For each threshold bin $i$ (we used $N_{bins} = 25$ threshold bins per functional, and hence $4 \cdot N_{bins} =100$ bins in total) we computed the probability $p_i$
that the Minkowski functional values for the string simulation come from the same
distribution as obtained from the pure noise maps. We then combined the
results of different bins into a $\chi^2$ value using the Fisher combined probability test
\begin{equation}
\chi^2 = -2 \displaystyle \sum_{i=1}^{4 N_{bins}} log(p_i) \, .
\end{equation}
We then computed the mean value ${\bar{\chi^2}}$ and the standard deviation 
$\Delta \chi^2$ of $\chi^2$. The criterion for significance of the difference
between string wake and noise maps is that the value of 
${\bar{\chi^2}} - \Delta \chi^2$ is larger than the representative significance
value for an analysis with $2( 4 N_{bins} )= 200$ degrees of freedom .
 
A comparison of the functionals of (wakes+noise) against pure noise is shown in 
Figures 5 to 8, which were done using a lattice size of $128^3$, each for a different
value of $G \mu$. In each set of graphs, the black curves represent the 
results for the pure noise maps (the mean values and standard deviations for
each threshold value are shown), and the orange curves (the less symmetric
- in the case of $V_1$ and $V_3$ - or less antisymmetric - in the
case of $V_0$ and $V_2$ curves) represent what is obtained
for maps containing both strings and noise. What is shown are the
Minkowski functionals for a particular cosmic string wake simulation.
The mean value of $\chi^2$ was 
found to be greater than $10^3$ (the maximal value our numerics
could handle) for $G \mu \geq 6.7\times 10^{-8}$, showing that
for these values of $G \mu$ string maps can indeed be distinguished
from pure noise maps.  For $G \mu = 4.2\times10^{-8}$ the $\chi^2$ drops down to $223\pm 101$, which indicates a good chance that any one set of wakes+noise will be indistinguishable from the noise at this value of $G \mu$. Hence from this preliminary analysis we conclude 
that the minimum string tension $G\mu$ must be roughly $5\times10^{-8}$ for 
the Minkowski functional method to be able to detect the cosmic string signals
which are embedded in the noise. 

\begin{center}
\begin{figure*}
\begin{center}
\label{fig:wakesAndnoise0}
  \caption{Minkowski functionals of a combined map (wakes+noise) with 
  $G\mu =1.7\times 10^{-7}$ along with the corresponding functionals for a 
  pure noise map, averaged over 100 simulations each using a lattice of $128^3$ pixels. 
  The black curves are for the pure noise maps, the orange ones for the strings+noise
  maps. }
  \centering
    \includegraphics[scale=1]{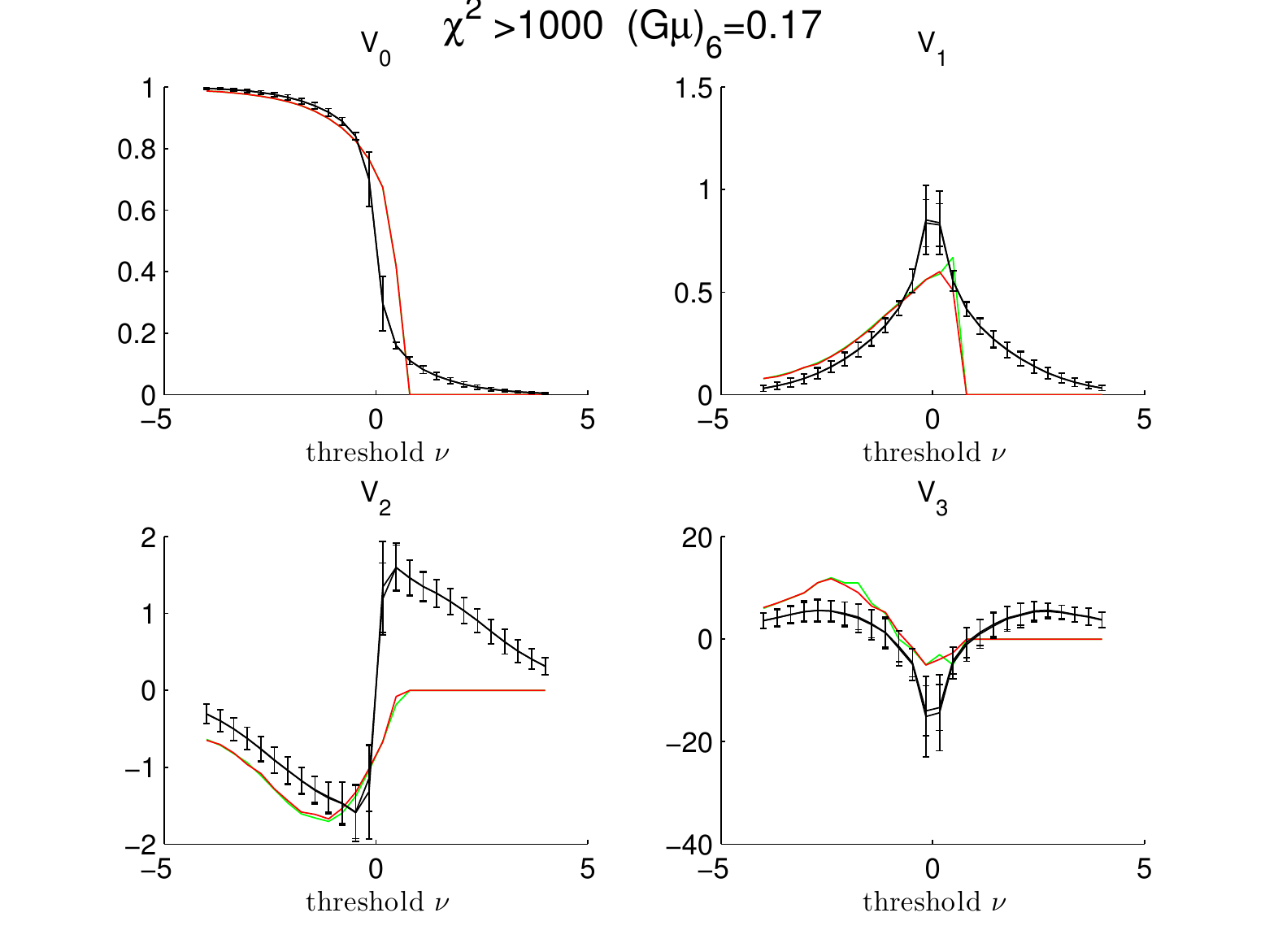}
\end{center}
\end{figure*}
\end{center}

\begin{center}
\begin{figure*}
\begin{center}
\label{fig:wakesAndnoise1}
  \caption{Minkowski functionals of a combined map (wakes+noise) with 
  $G\mu =1.07\times 10^{-7}$ along with the corresponding functionals for a 
  pure noise map, averaged over 100 simulations each using a lattice of $128^3$ pixels.}
  \centering
    \includegraphics[scale=1]{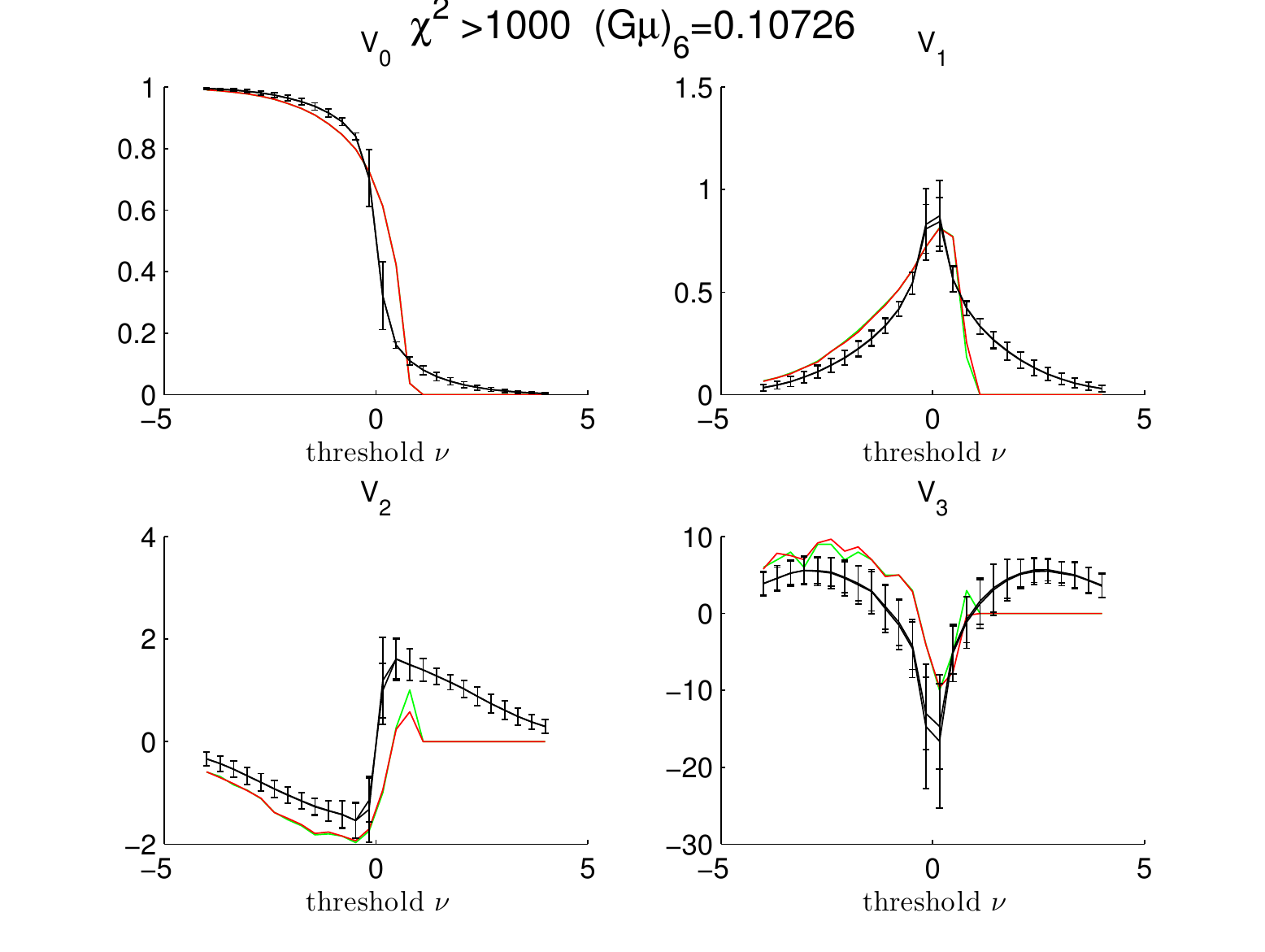}
\end{center}
\end{figure*}
\end{center}

\begin{center}
\begin{figure*}
\begin{center}
\label{fig:wakesAndnoise2}
  \caption{Minkowski functionals of a combined map (wakes+noise) with 
  $G\mu =6.7\times 10^{-8}$ along with the corresponding functionals for a pure noise map, averaged over 100 simulations each using a lattice of $128^3$ pixels.}
  \centering
    \includegraphics[scale=1]{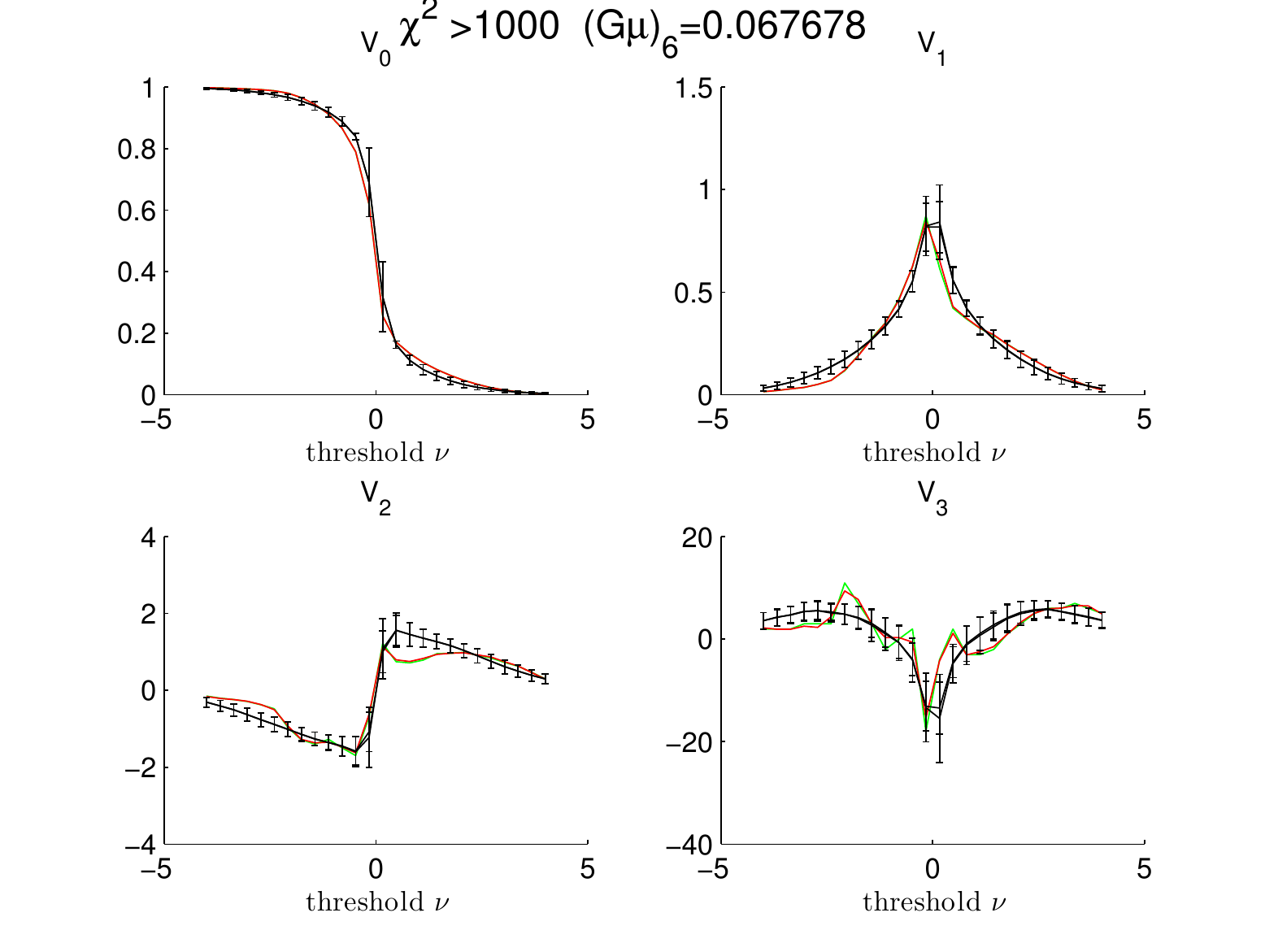}
\end{center}
\end{figure*}
\end{center}

\begin{center}
\begin{figure*}
\begin{center}
\label{fig:wakesAndnoise3}
  \caption{Minkowski functionals of a combined map (wakes+noise) with 
  $G\mu =4.3\times 10^{-8}$ along with the corresponding functionals for a pure 
  noise map, averaged over 100 simulations each using a lattice of $128^3$ pixels.}
  \centering
    \includegraphics[scale=1]{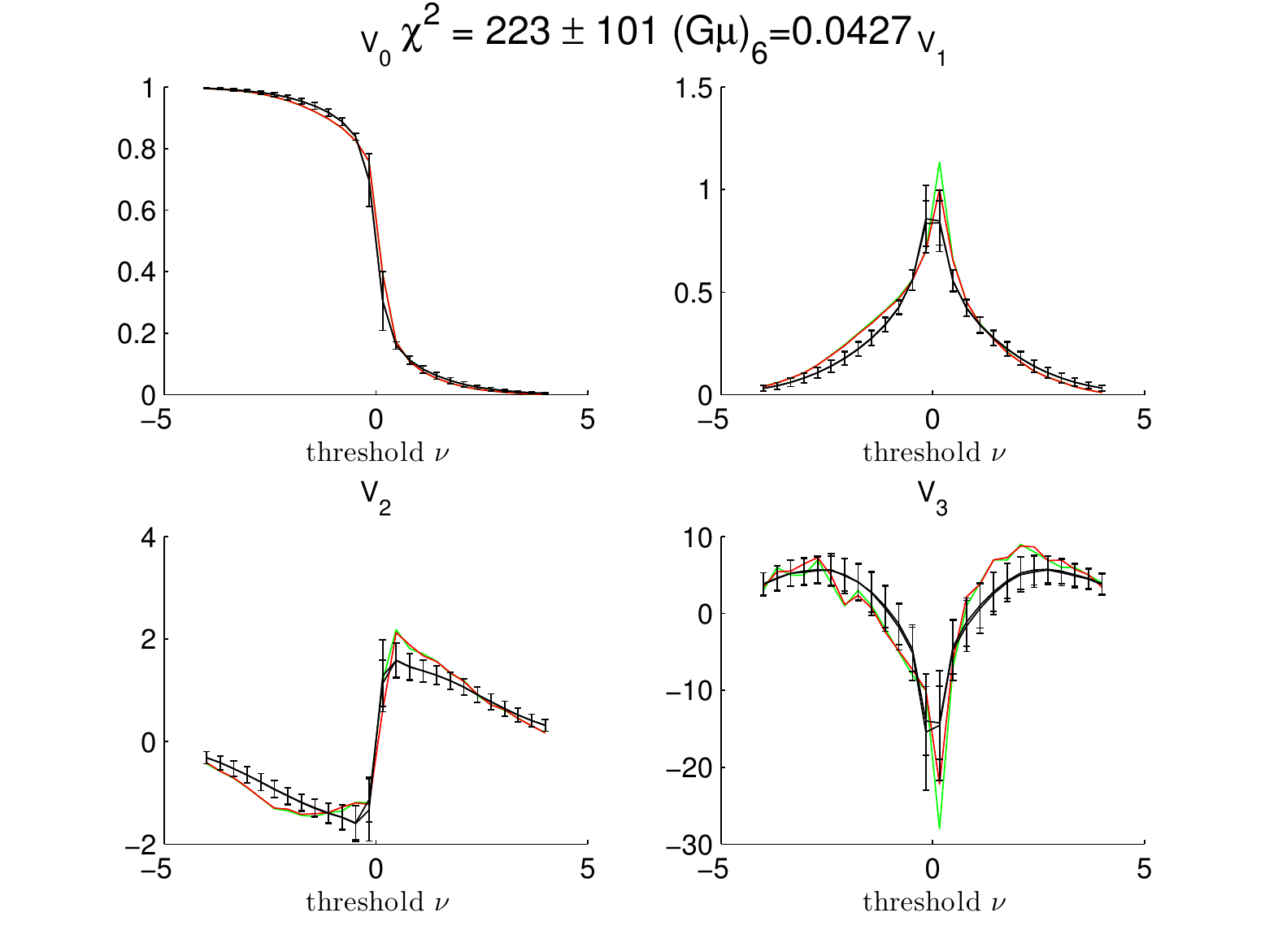}
\end{center}
\end{figure*}
\end{center}

Note the $V_i$ values for the cosmic string simulations in Figure 5 vanish to the
level of the numerical accuracy above a certain value of the threshold, while those for
noise alone do not. The reason this happens is that in the presence of string wakes, a
fixed threshold value corresponds to a different brightness temperature than it would
in the absence of strings.

Given the limited accuracy of the analysis, we know that the Minkowski functionals at sufficiently
close threshold values are not independent. We must thus consider the possibility that we
might have oversampled and obtained an artificially large value of $\chi^2$.  Put another
way, since the temperature resolution $\delta T$ of our maps is non-zero, we must account 
for this in our choice for the threshold bin resolution $\delta \nu$. We can relate $\delta \nu$ 
and $\delta T$ by
\begin{equation}
\delta \nu = \frac{\delta T}{\sigma_T} \, .
\label{eq:deltanu}
\end{equation}
The temperature resolution can be estimated to first order by 
\begin{equation} 
\delta T =  \delta x \cdot \nabla T \, , 
\end{equation}
where $\delta x$ is the spatial resolution and $\nabla T$ is the gradient of the 
background noise. Since the resolution of the maps is the worst in the redshift 
direction, we focus on resolution in the $z$-direction. The relevant quantity then 
is the spatial resolution $\delta l$ corresponding to the redshift resolution 
$\delta z$, which is given by
\begin{equation}
\delta l (\delta z, z )\sim c  (\delta z) t_0 z^{-3/2} \, .
\end{equation}
 To then calculate the gradient $\nabla T$, we use the fact that that the matter power 
spectrum is dominated by modes around $k\sim1/\lambda_{eq}$, and hence this 
sets the length scale of the problem. We can then estimate the gradient by
\begin{equation}
T= \frac{\bar{T}}{ \lambda_{eq}} \, ,
\end{equation}
where $\lambda_{eq}$ is
\begin{equation}
\lambda_{eq}  \sim c t_{eq} z_{eq}^{-1/2} \, .
\end{equation}

Putting this all together we find
\begin{equation}
\delta T \sim \bar{T} \left( \frac{\delta z}{ z} \right) \left( \frac{z_{eq}}{  z}\right)^{1/2}  
\sim 0.05 \mbox{mK} \, ,
\end{equation}
where for the second equality we take $z=50$. Using equation \ref{eq:deltanu}, 
this corresponds to $\delta \nu = 0.036$. We can then translate that into an upper bound 
on the number of threshold bins we should use to span 8 standard deviations 
($-4 \sigma$ to $4 \sigma$) and find
\begin{equation}
N_{bins} = \frac{8}{\delta \nu} \sim 200 \, .
\end{equation}
Hence our choice of 25 threshold bins is well within the limit from oversampling.

Another way to see that the bins can be treated as independent is to consider the correlation matrix for each functional. The correlation matrix is defined by
\be
Corr_{ij} = \frac{Cov_{ij}}{\sigma_i \sigma_j}  = \frac{\langle ( X_i - \mu_i )( X_j - \mu_j ) \rangle}{\sigma_i \sigma_j}
\ee
where the $X_i$ are defined by the wake+noise signal for the $i'th$ bin. Hence $i$ runs from $1$ to $25$, each $X_i$ is a random variable that we have sampled 100 times (once for each trial of the simulation), and $\langle \rangle$ denotes expectation value. From each $X_i$ we define a mean $\langle X_i \rangle = \mu_i$ and a standard deviation $\sigma_i$. From this we can calculate the correlation in the wake signal between the bins for each functional, as is plotted in Figure 9. For each functional, the correlation matrix is concentrated in a narrow strip along the diagonal with a width  at half max of 2-3 bins. Given this diagonal structure, we are safe from any hidden correlation that would invalidate our use of the Fisher combined probability test.

\begin{center}
\begin{figure*}
\begin{center}
\label{fig:CorrMatrix}
  \caption{Correlation of wake signal between bins for each functional, using data from 100 realizations.}
  \centering
    \includegraphics[scale=0.5]{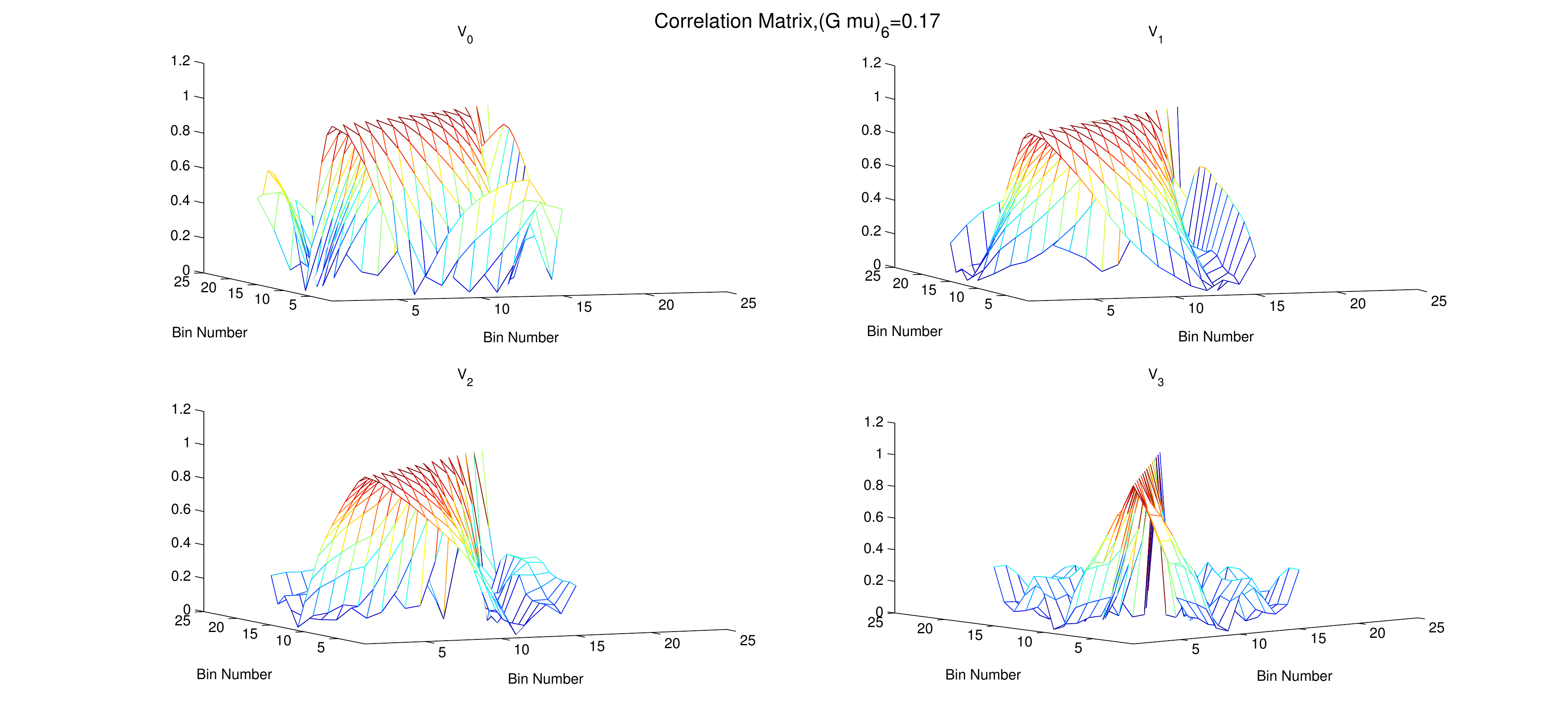}
\end{center}
\end{figure*}
\end{center}

\section{Summary and Outlook}

This purpose of this investigation has been to both demonstrate the power of 
Minkowski Functionals for detecting non-Gaussian behaviour, and to investigate 
the topology of 21cm distributions in cosmological models with a scaling
distribution of cosmic string wakes. We generated  3-D maps of the 
21cm brightness temperature in pure cosmic string models, in models
with only background noise from mini-halos, and in maps in which both
sources of 21cm fluctuations were present, using methods similar to those 
employed for constructing 2-D CMB temperature by \cite{Danos0}. These maps 
were then be investigated with Minkowski functionals, using the software 
`\emph{Minkowski3}'  made available by Buchert \cite{Buchert}. For large
values of the string tension, the non-Gaussian signatures of string wakes are
clearly visible in the Minkowski functionals. We studied for which range of
values of the string tension the difference between the strings + noise
versus the pure noise maps as seen in the Minkowski functionals are
statictically significant. 
A large difference was found between the functionals of the string-induced 
brightness temperature map and the corresponding functionals of background noise, 
with maps of strings embedded in noise found to be differentiable from maps of pure 
noise when $G \mu > 5\times 10^{-8}$. 

The analysis presented in this report serves as a proof of concept that the 
Minkowski functionals can be used as a statistical tool to search for cosmic
string signals. To extend this preliminary analysis would mean performing 
large scale numerical simulations, as new features may present themselves as 
we study the 21cm maps with greater resolution.

\begin{acknowledgments} 
 
We thank Thomas Buchert, for permitting the use of his software, and Rebecca Danos 
for permission to use Figure 1 which is taken from \cite{Danos2}. We also wish
to thank an anonymous referee for many suggestions on how to improve this
work. This research has been 
supported in part by an NSERC Discovery Grant, by funds from the CRC, and by a 
Killam Research Fellowship. 
 
\end{acknowledgments}


\begin{thebibliography}{99} 

\bibitem{Rachel}
R.~Jeannerot,
  ``A Supersymmetric SO(10) Model with Inflation and Cosmic Strings,''
  Phys.\ Rev.\  D {\bf 53}, 5426 (1996)
  [arXiv:hep-ph/9509365];\\
R.~Jeannerot, J.~Rocher and M.~Sakellariadou,
  ``How generic is cosmic string formation in SUSY GUTs,''
  Phys.\ Rev.\  D {\bf 68}, 103514 (2003)
  [arXiv:hep-ph/0308134].

\bibitem{Tye}
S.~Sarangi and S.~H.~H.~Tye,
  ``Cosmic string production towards the end of brane inflation,''
  Phys.\ Lett.\  B {\bf 536}, 185 (2002)
  [arXiv:hep-th/0204074];\\
E.~J.~Copeland, R.~C.~Myers and J.~Polchinski,
  ``Cosmic F- and D-strings,''
  JHEP {\bf 0406}, 013 (2004)
  [arXiv:hep-th/0312067].

\bibitem{SGC}
R.~H.~Brandenberger and C.~Vafa,
  ``Superstrings in the Early Universe,''
  Nucl.\ Phys.\  B {\bf 316}, 391 (1989);\\
 A.~Nayeri, R.~H.~Brandenberger and C.~Vafa, 
  ``Producing a scale-invariant spectrum of perturbations in a Hagedorn phase 
  of string cosmology,''
 Phys.\ Rev.\ Lett.\  {\bf 97}, 021302 (2006)   [arXiv:hep-th/0511140];\\
 R.~H.~Brandenberger, A.~Nayeri, S.~P.~Patil and C.~Vafa,
  ``String gas cosmology and structure formation,''
 Int.\ J.\ Mod.\ Phys.\ A {\bf 22}, 3621 (2007)
  [hep-th/0608121];\\
  R.~H.~Brandenberger,
  ``String Gas Cosmology,''
  arXiv:0808.0746 [hep-th].

\bibitem{ShelVil}
A. Vilenkin and E.P.S. Shellard, \textit{Cosmic Strings and other
Topological Defects} (Cambridge Univ. Press, Cambridge, 1994).

\bibitem{HK}
M.~B.~Hindmarsh and T.~W.~B.~Kibble,
  ``Cosmic strings,''
  Rept.\ Prog.\ Phys.\  {\bf 58}, 477 (1995)
  [arXiv:hep-ph/9411342].

\bibitem{RHBrev}
R.~H.~Brandenberger,
  ``Topological defects and structure formation,''
  Int.\ J.\ Mod.\ Phys.\ A {\bf 9}, 2117 (1994)
  [arXiv:astro-ph/9310041].

\bibitem{CSearly}
 Y.~B.~Zeldovich,
  ``Cosmological fluctuations produced near a singularity,''
  Mon.\ Not.\ Roy.\ Astron.\ Soc.\  {\bf 192}, 663 (1980);\\
 A.~Vilenkin,
  ``Cosmological Density Fluctuations Produced By Vacuum Strings,''
  Phys.\ Rev.\ Lett.\  {\bf 46}, 1169 (1981)
  [Erratum-ibid.\  {\bf 46}, 1496 (1981)];\\
N.~Turok and R.~H.~Brandenberger,
  ``Cosmic Strings And The Formation Of Galaxies And Clusters Of Galaxies,''
  Phys.\ Rev.\ D {\bf 33}, 2175 (1986);\\
H. Sato, ``Galaxy Formation by Cosmic Strings,''
  Prog. Theor. Phys.\  {\bf 75}, 1342 (1986);\\
A. Stebbins, ``Cosmic Strings and Cold Matter'',
  Ap. J. (Lett.) {\bf 303}, L21 (1986).

\bibitem{Pagano}
M.~Pagano and R.~Brandenberger,
  ``The 21cm Signature of a Cosmic String Loop,''
  JCAP {\bf 1205}, 014 (2012)
  [arXiv:1201.5695 [astro-ph.CO]].

\bibitem{CSnum}
A.~Albrecht and N.~Turok,
  ``Evolution Of Cosmic Strings,''
  Phys.\ Rev.\ Lett.\  {\bf 54}, 1868 (1985);\\
D.~P.~Bennett and F.~R.~Bouchet,
  ``Evidence For A Scaling Solution In Cosmic String Evolution,''
  Phys.\ Rev.\ Lett.\  {\bf 60}, 257 (1988);\\
B.~Allen and E.~P.~S.~Shellard,
  ``Cosmic String Evolution: A Numerical Simulation,''
  Phys.\ Rev.\ Lett.\  {\bf 64}, 119 (1990);\\
C.~Ringeval, M.~Sakellariadou and F.~Bouchet,
  ``Cosmological evolution of cosmic string loops,''
  JCAP {\bf 0702}, 023 (2007)
  [arXiv:astro-ph/0511646];\\
  A.~A.~Fraisse, C.~Ringeval, D.~N.~Spergel and F.~R.~Bouchet,
  ``Small-Angle CMB Temperature Anisotropies Induced by Cosmic Strings,''
  Phys.\ Rev.\ D {\bf 78}, 043535 (2008)
  [arXiv:0708.1162 [astro-ph]];\\
C.~J.~A.~P.~Martins and E.~P.~S.~Shellard,
  ``Fractal properties and small-scale structure of cosmic string networks,''
  Phys.\ Rev.\ D {\bf 73}, 043515 (2006)
  [astro-ph/0511792];\\
V.~Vanchurin, K.~D.~Olum and A.~Vilenkin,
  ``Scaling of cosmic string loops,''
  Phys.\ Rev.\  D {\bf 74}, 063527 (2006)
  [arXiv:gr-qc/0511159];\\
 J.~J.~Blanco-Pillado, K.~D.~Olum and B.~Shlaer,
  ``Large parallel cosmic string simulations: New results on loop production,''
  Phys.\ Rev.\ D {\bf 83}, 083514 (2011)
  [arXiv:1101.5173 [astro-ph.CO]];\\
  C.~Ringeval and F.~R.~Bouchet,
  ``All sky CMB map from cosmic strings integrated Sachs-Wolfe effect,''
  arXiv:1204.5041 [astro-ph.CO].
  
\bibitem{deficit}
A.~Vilenkin,
  ``Gravitational Field Of Vacuum Domain Walls And Strings,''
  Phys.\ Rev.\  D {\bf 23}, 852 (1981);\\
R.~Gregory,
  ``Gravitational Stability of Local Strings,''
  Phys.\ Rev.\ Lett.\  {\bf 59}, 740 (1987).

\bibitem{wake}
J.~Silk and A.~Vilenkin,
  ``Cosmic Strings And Galaxy Formation,''
  Phys.\ Rev.\ Lett.\  {\bf 53}, 1700 (1984).

\bibitem{wakegrowth}
 L.~Perivolaropoulos, R.~H.~Brandenberger and A.~Stebbins,
  ``Dissipationless Clustering Of Neutrinos In Cosmic String Induced Wakes,''
  Phys.\ Rev.\  D {\bf 41}, 1764 (1990);\\
  R.~H.~Brandenberger, L.~Perivolaropoulos and A.~Stebbins,
  ``Cosmic Strings, Hot Dark Matter and the Large Scale Structure of the 
  Universe,''
  Int.\ J.\ Mod.\ Phys.\  A {\bf 5}, 1633 (1990).

\bibitem{Dvorkin}
  C.~Dvorkin, M.~Wyman and W.~Hu,
  ``Cosmic String constraints from WMAP and the South Pole Telescope,''
  Phys.\ Rev.\ D {\bf 84}, 123519 (2011)
  [arXiv:1109.4947 [astro-ph.CO]].

\bibitem{CSbound}
L.~Pogosian, S.~H.~H.~Tye, I.~Wasserman and M.~Wyman,
  ``Observational constraints on cosmic string production during brane
  inflation,''
  Phys.\ Rev.\  D {\bf 68}, 023506 (2003)
  [Erratum-ibid.\  D {\bf 73}, 089904 (2006)]
  [arXiv:hep-th/0304188];\\
M.~Wyman, L.~Pogosian and I.~Wasserman,
  ``Bounds on cosmic strings from WMAP and SDSS,''
  Phys.\ Rev.\  D {\bf 72}, 023513 (2005)
  [Erratum-ibid.\  D {\bf 73}, 089905 (2006)]
  [arXiv:astro-ph/0503364];\\
A.~A.~Fraisse,
  ``Limits on Defects Formation and Hybrid Inflationary Models with
  Three-Year WMAP Observations,''
  JCAP {\bf 0703}, 008 (2007)
  [arXiv:astro-ph/0603589];\\
U.~Seljak, A.~Slosar and P.~McDonald,
  ``Cosmological parameters from combining the Lyman-alpha forest with CMB,
  galaxy clustering and SN constraints,''
  JCAP {\bf 0610}, 014 (2006)
  [arXiv:astro-ph/0604335];\\
  R.~A.~Battye, B.~Garbrecht and A.~Moss,
  ``Constraints on supersymmetric models of hybrid inflation,''
  JCAP {\bf 0609}, 007 (2006)
  [arXiv:astro-ph/0607339];\\
R.~A.~Battye, B.~Garbrecht, A.~Moss and H.~Stoica,
  ``Constraints on Brane Inflation and Cosmic Strings,''
  JCAP {\bf 0801}, 020 (2008)
  [arXiv:0710.1541 [astro-ph]];\\
N.~Bevis, M.~Hindmarsh, M.~Kunz and J.~Urrestilla,
  ``CMB power spectrum contribution from cosmic strings using  field-evolution
  simulations of the Abelian Higgs model,''
  Phys.\ Rev.\  D {\bf 75}, 065015 (2007)
  [arXiv:astro-ph/0605018];\\
N.~Bevis, M.~Hindmarsh, M.~Kunz and J.~Urrestilla,
  ``Fitting CMB data with cosmic strings and inflation,''
 Phys.\ Rev.\ Lett.\  {\bf 100}, 021301 (2008)
  [astro-ph/0702223 [ASTRO-PH]];\\
R.~Battye and A.~Moss,
  ``Updated constraints on the cosmic string tension,''
  Phys.\ Rev.\ D {\bf 82}, 023521 (2010)
  [arXiv:1005.0479 [astro-ph.CO]].

\bibitem{Guth}
A. Guth, 
 ``The Inflationary Universe: A Possible Solution To The Horizon And Flatness
 Problems,''
  Phys.\ Rev.\  D {\bf 23}, 347 (1981).

\bibitem{Mukh}
V. Mukhanov and G. Chibisov,
  ``Quantum Fluctuation And Nonsingular Universe. (In Russian),''
  JETP Lett.\  {\bf 33}, 532 (1981)
  [Pisma Zh.\ Eksp.\ Teor.\ Fiz.\  {\bf 33}, 549 (1981)].

\bibitem{LSS}
 M. Rees,
 ``Baryon concentrations in string wakes at $z \geq 200$:
 implications for galaxy formation and large-scale structure,"
 Mon. Not. R. astr. Soc. {\bf{222}}, 27p (1986);\\
T.~Vachaspati,
  ``Cosmic Strings and the Large-Scale Structure of the Universe,''
  Phys.\ Rev.\ Lett.\  {\bf 57}, 1655 (1986);\\
A.~Stebbins, S.~Veeraraghavan, R.~H.~Brandenberger, J.~Silk and N.~Turok,
  ``Cosmic String Wakes,''
  Astrophys.\ J.\  {\bf 322}, 1 (1987)'\\
J.~C.~Charlton,
  ``Cosmic String Wakes and Large Scale Structure,''
  Astrophys.\ J.\  {\bf 325}, 52 (1988);\\
T.~Hara and S.~Miyoshi,
  ``Formation of the First Systems in the Wakes of Moving Cosmic Strings,''
  Prog.\ Theor.\ Phys.\  {\bf 77}, 1152 (1987);\\
T.~Hara and S.~Miyoshi,
  ``Large Scale Structures and Streaming Velocities Due to Open Cosmic 
  Strings,''
  Prog.\ Theor.\ Phys.\  {\bf 81}, 1187 (1989).

\bibitem{Danos1}
 R.~J.~Danos, R.~H.~Brandenberger and G.~Holder,
  ``A Signature of Cosmic Strings Wakes in the CMB Polarization,''
  Phys.\ Rev.\  D {\bf 82}, 023513 (2010)
  [arXiv:1003.0905 [astro-ph.CO]].

\bibitem{Danos2}
 R.~H.~Brandenberger, R.~J.~Danos, O.~F.~Hernandez and G.~P.~Holder,
  ``The 21 cm Signature of Cosmic String Wakes,''
  JCAP {\bf 1012}, 028 (2010)
  [arXiv:1006.2514 [astro-ph.CO]].

\bibitem{Wangyi}
  O.~F.~Hernandez, Y.~Wang, R.~Brandenberger and J.~Fong,
  ``Angular 21 cm Power Spectrum of a Scaling Distribution of Cosmic String
  Wakes,''
 JCAP {\bf 1108}, 014 (2011)
  [arXiv:1104.3337 [astro-ph.CO]].

\bibitem{Hadwiger}
H. Hadwiger, {\it Vorlesungen \"uber Inhalt, Oberfl\"ache und Isoperimetrie} 
(Springer, Berlin, 1957).

\bibitem{superclusters}
K.~R.~Mecke, T.~Buchert, H.~Wagner,
  ``Robust morphological measures for large scale structure in the universe,''
  Astron.\ Astrophys.\  {\bf 288}, 697-704 (1994).
  [astro-ph/9312028];\\
J.~Schmalzing and T.~Buchert,
  ``Beyond genus statistics: a unifying approach to the morphology of cosmic
  structure,''
  Astrophys.\ J.\  {\bf 482}, L1 (1997)
  [arXiv:astro-ph/9702130];\\
J.~Schmalzing, T.~Buchert, A.~L.~Melott, V.~Sahni, B.~S.~Sathyaprakash and S.~F.~Shandarin,
  ``Disentangling the cosmic web I: morphology of isodensity contours,''
  Astrophys.\ J.\  {\bf 526}, 568 (1999)
  [arXiv:astro-ph/9904384].
  
\bibitem{CMB}
D.~Novikov, H.~A.~Feldman and S.~F.~Shandarin,
  ``Minkowski functionals and cluster analysis for CMB maps,''
  Int.\ J.\ Mod.\ Phys.\  D {\bf 8}, 291 (1999)
  [arXiv:astro-ph/9809238];\\
  C.~Hikage, E.~Komatsu and T.~Matsubara,
  ``Primordial Non-Gaussianity and Analytical Formula for Minkowski Functionals
  of the Cosmic Microwave Background and Large-scale Structure,''
  Astrophys.\ J.\  {\bf 653}, 11 (2006)
  [arXiv:astro-ph/0607284];\\
S.~Winitzki and A.~Kosowsky,
  ``Minkowski functional description of microwave background Gaussianity,''
  New Astron.\  {\bf 3}, 75 (1998)
  [arXiv:astro-ph/9710164].

\bibitem{CSMink}
D.~Mitsouras, R.~H.~Brandenberger and P.~Hickson,
  ``Topological Statistics and the LMT Galaxy Redshift Survey,''
  arXiv:astro-ph/9806360;\\
H.~Trac, D.~Mitsouras, P.~Hickson and R.~H.~Brandenberger,
  ``Topology of the Las Campanas Redshift Survey,''
  Mon.\ Not.\ Roy.\ Astron.\ Soc.\  {\bf 330}, 531 (2002)
  [arXiv:astro-ph/0007125].

 \bibitem{ACT}
A.~Kosowsky  [the ACT Collaboration],
  ``The Atacama Cosmology Telescope Project: A Progress Report,''
  New Astron.\ Rev.\  {\bf 50}, 969 (2006)
  [arXiv:astro-ph/0608549].
  
\bibitem{SPT}
 J.~E.~Ruhl {\it et al.}  [The SPT Collaboration],
  ``The South Pole Telescope,''
  Proc.\ SPIE Int.\ Soc.\ Opt.\ Eng.\  {\bf 5498}, 11 (2004)
  [arXiv:astro-ph/0411122];\\
  J.~E.~Carlstrom {\it et al.},
  ``The 10 Meter South Pole Telescope,''
  Publ.\ Astron.\ Soc.\ Pac.\  {\bf 123}, 568 (2011)
  [arXiv:0907.4445 [astro-ph.IM]].

\bibitem{CMBandACT}
J. Urrestilla, N. Bevis, M. Hindmarsh, and M. Kunz,
``Cosmic string parameter constraints and model analysis using small scale Cosmic Microwave Background data,''
  JCAP {\bf 1112}, 021 (2011)
  [arXiv:1108.2730 [astro-ph.CO]].

\bibitem{Amsel}
S.~Amsel, J.~Berger and R.~H.~Brandenberger,
  ``Detecting Cosmic Strings in the CMB with the Canny Algorithm,''
  JCAP {\bf 0804}, 015 (2008)
  [arXiv:0709.0982 [astro-ph]].

\bibitem{Stewart}
A.~Stewart and R.~Brandenberger,
  ``Edge Detection, Cosmic Strings and the South Pole Telescope,''
 JCAP {\bf 0902}, 009 (2009)
  [arXiv:0809.0865 [astro-ph]].
  
\bibitem{Danos0} 
  R.~J.~Danos and R.~H.~Brandenberger,
  ``Canny Algorithm, Cosmic Strings and the Cosmic Microwave Background,''
  Int.\ J.\ Mod.\ Phys.\ D {\bf 19}, 183 (2010)
  [arXiv:0811.2004 [astro-ph]].

\bibitem{KS}
N.~Kaiser and A.~Stebbins,
  ``Microwave Anisotropy Due To Cosmic Strings,''
  Nature {\bf 310}, 391 (1984).

\bibitem{Ravi}
M.~S.~Movahed, B.~Javanmardi and R.~K.~Sheth,
   ``Peak-peak correlations in the cosmic background radiation from cosmic 
strings,''
   arXiv:1212.0964 [astro-ph.CO].

\bibitem{CSGR}
T.~Vachaspati and A.~Vilenkin,
  ``Gravitational Radiation from Cosmic Strings,''
  Phys.\ Rev.\ D {\bf 31}, 3052 (1985);\\
R. L. Davis, 
``Nucleosynthesis Problems for String Models of Galaxy Formation",
Phys. Lett. {\bf B 161}, 285 (1985);\\
R.~H.~Brandenberger, A.~Albrecht and N.~Turok,
  ``Gravitational Radiation From Cosmic Strings And The Microwave Background,''
  Nucl.\ Phys.\ B {\bf 277}, 605 (1986).
  
\bibitem{pulsar}
D.~R.~Stinebring, M.~F.~Ryba, J.~H.~Taylor and R.~W.~Romani,
   ``The Cosmic Gravitational Wave Background: Limits From Millisecond 
Pulsar Timing,''
   Phys.\ Rev.\ Lett.\  {\bf 65}, 285 (1990);\\
F.~R.~Bouchet and D.~P.~Bennett,
   ``Does The Millisecond Pulsar Constrain Cosmic Strings?,''
   Phys.\ Rev.\ D {\bf 41}, 720 (1990);\\
  R.~R.~Caldwell and B.~Allen,
   ``Cosmological constraints on cosmic string gravitational radiation,''
   Phys.\ Rev.\ D {\bf 45}, 3447 (1992);\\
   S.~A.~Sanidas, R.~A.~Battye and B.~W.~Stappers,
   ``Projected constraints on the cosmic (super)string tension with future 
gravitational wave detection experiments,''
   arXiv:1211.5042 [astro-ph.CO];\\
S.~Kuroyanagi, K.~Miyamoto, T.~Sekiguchi, K.~Takahashi and J.~Silk,
   ``Forecast constraints on cosmic strings from future CMB, pulsar timing 
and gravitational wave direct detection experiments,''
   arXiv:1210.2829 [astro-ph.CO].
  
\bibitem{Abbott:2007kv} 
  B.~Abbott {\it et al.}  [LIGO Scientific Collaboration],
  Rept.\ Prog.\ Phys.\  {\bf 72}, 076901 (2009)
  [arXiv:0711.3041 [gr-qc]].

\bibitem{SKA}
http://www.skatelescope.org/

\bibitem{E-ELT}
http://www.eso.org/sci/facilities/eelt/

\bibitem{lofar}
See http://www.lofar.org/ .

\bibitem{Holwerda:2011kd} 
  B.~W.~Holwerda, S.~--L.~Blyth, A.~J.~Baker and t.~L.~team,
  ``Looking At the Distant Universe with the MeerKAT Array (LADUMA),''
  arXiv:1109.5605 [astro-ph.CO].

\bibitem{Periv}
L.~Perivolaropoulos,
  ``COBE versus cosmic strings: An Analytical model,''
  Phys.\ Lett.\  B {\bf 298}, 305 (1993)
  [arXiv:hep-ph/9208247];\\
L.~Perivolaropoulos,
  ``Statistics of microwave fluctuations induced by topological defects,''
  Phys.\ Rev.\  D {\bf 48}, 1530 (1993)
  [arXiv:hep-ph/9212228].

\bibitem{Furl}
S.~Furlanetto, S.~P.~Oh and F.~Briggs,
  ``Cosmology at Low Frequencies: The 21 cm Transition and the High-Redshift
  Universe,''
  Phys.\ Rept.\  {\bf 433}, 181 (2006)
  [arXiv:astro-ph/0608032].

\bibitem{diffuse}
O.~F.~Hernandez and R.~H.~Brandenberger,
  ``The 21 cm Signature of Shock Heated and Diffuse Cosmic String Wakes,''
  JCAP, in press (2012), [arXiv:1203.2307 [astro-ph.CO]].

\bibitem{sornborger}
A.~Sornborger, R.~H.~Brandenberger, B.~Fryxell and K.~Olson,
  ``The structure of cosmic string wakes,''
  Astrophys.\ J.\  {\bf 482}, 22 (1997)
  [arXiv:astro-ph/9608020].

\bibitem{shapiroAnalytical}
 Paul R. Shapiro and Kyungjin Ahn and Marcelo A. Alvarez and Ilian T. Iliev and Hugo Martel and Dongsu Ryu,
  ``The 21 cm Background from the Cosmic Dark Ages: Minihalos and the Intergalactic Medium before Reionization''
The Astrophysical Journal,  {\bf 646}, 2,
http://stacks.iop.org/0004-637X/646/i=2/a=681,
2006

 \bibitem{arXiv:1005.2502} 
  I.~T.~Iliev, K.~Ahn, J.~Koda, P.~R.~Shapiro and U.~-L.~Pen,
  ``Cosmic Structure Formation at High Redshift,''
  arXiv:1005.2502 [astro-ph.CO].

\bibitem{Press:1973iz}
  W.~H.~Press and P.~Schechter,
  ``Formation of galaxies and clusters of galaxies by selfsimilar gravitational
  condensation,''
  Astrophys.\ J.\  {\bf 187}, 425 (1974).

\bibitem{Iliev:2002gj} 
  I.~T.~Iliev, P.~R.~Shapiro, A.~Ferrara and H.~Martel,
  ``On the direct detectability of the cosmic dark ages: 21-cm emission from minihalos,''
  Astrophys.\ J.\  {\bf 572}, 123 (2002)
  [astro-ph/0202410].

\bibitem{Peebles}
P. J. E. Peebles, {\it The Large-Scale Structure of the Universe}
(Princeton Univ. Press, Princeton, 1980).

\bibitem{Minkowski}
H. Minkowski, Mathematische Annalen {\bf 57}, 447 (1903).

\bibitem{Schmalzing1}
 J.~Schmalzing and K.~M.~Gorski,
  ``Minkowski functionals used in the morphological analysis of cosmic
  microwave background anisotropy maps,''
  arXiv:astro-ph/9710185.
  
\bibitem{Schmalzing2}
J.~Schmalzing, M.~Kerscher and T.~Buchert,
  ``Minkowski functionals in cosmology,''
  arXiv:astro-ph/9508154.

\bibitem{Tomita}
H. Tomita, in ``Formation, Dynamics and Statistics of Patterns" (Vol. 1) , ed. by
K. Kawasaki, M. Suzuki and A. Onuki (World Scientific, Singapore, 1990).

\bibitem{minkowskicalculation}
 J.~Schmalzing and T.~Buchert,
  ``Beyond genus statistics: a unifying approach to the morphology of cosmic
  structure,''
  Astrophys.\ J.\  {\bf 482}, L1 (1997)
  [arXiv:astro-ph/9702130].
  
\bibitem{Buchert}
T. Buchert, `Minkowski3', 1997
http://www.cosmunix.de/buchert\_software.htm .

    
\end{thebibliography}
\end{document}